\documentclass{article}
\usepackage{iclr2027_conference,times}

\newsavebox{\pwbtab}

\usepackage{amsmath,amsfonts,bm}

\def\eqref#1{equation~\ref{#1}}

\def\1{\bm{1}}

\DeclareMathAlphabet{\mathsfit}{\encodingdefault}{\sfdefault}{m}{sl}
\SetMathAlphabet{\mathsfit}{bold}{\encodingdefault}{\sfdefault}{bx}{n}

\usepackage{hyperref}
\usepackage{url}
\usepackage{graphicx}
\usepackage{booktabs}
\usepackage{amsmath}
\usepackage{amssymb}
\usepackage{xcolor}
\usepackage{multirow}
\usepackage{colortbl}
\usepackage{microtype}
\usepackage{enumitem}
\usepackage{float}
\usepackage{pifont}
\usepackage{wrapfig}
\definecolor{HeaderColor}{RGB}{226,230,236}
\definecolor{GoodColor}{RGB}{248,250,253}
\definecolor{ClosedColor}{RGB}{248,242,228}
\definecolor{SonnetTint}{RGB}{232,242,247}   
\definecolor{GPTTint}{RGB}{234,241,238}      
\definecolor{SonnetBand}{RGB}{197,219,232}   
\definecolor{GPTBand}{RGB}{200,221,213}      
\definecolor{BestTint}{RGB}{205,218,190}     
\newcommand{\bench}{\textsc{ProductWebBench}}
\newcommand{\wcs}{\mathrm{WCS}}

\title{\bench: Evaluating Website Continuity Beyond Task Completion}

\author{Jiajun Wu\textsuperscript{1}, Jian Yang\textsuperscript{1}, Yaxin Du\textsuperscript{2}, Shukai Liu\textsuperscript{1}, Linzheng Chai\textsuperscript{1}, Tuney Zheng\textsuperscript{2}\\ Wei Zhang\textsuperscript{1}, Haitao Yang\textsuperscript{3}, Yan Bai\textsuperscript{3}, Yihang Lou\textsuperscript{3}, Xianglong Liu\textsuperscript{1}, Bryan Dai\textsuperscript{2}\\[4pt] \textsuperscript{1}Beihang University\\ \textsuperscript{2}IQuest Research\\ \textsuperscript{3}Huawei Technologies Ltd.}

\iclrfinalcopy

\begin{document}
\maketitle

\begin{abstract}
Frontend coding agents are usually graded on one question: did the requested
target get built. Real engineering changes carry a second obligation, because the
rest of the site has to keep working. We call that second obligation
\emph{website continuity}, and introduce \bench{}, a benchmark that scores both on
living frontend repositories: 320 Change tasks over existing code
and 80 long-horizon Build tasks. Every task freezes a repository snapshot, staged
requirements, browser evidence and deterministic checks before evaluation, and
enters only once a reference passes every check and plausible-wrong ones fail.
A simulated user reveals the requirements one stage at a time, and every
check cleared earlier stays armed for the rest of the run. Across 8,737 runs from
13 models the two come apart: of the Change runs passing every
requested-work check, \textbf{28.2\%} still fail a continuity check -- a blank
route, horizontal overflow, console errors -- and \textbf{10.8\%} push a browser
state they had already passed back to failing. Because the same checks also
run on the untouched repository, each regression carries exact provenance, not a
post-hoc label: on the 586 runs where both readings exist, \textbf{18.3\%} are
caught breaking a verified state. Interaction is a
further blind spot, failing in \textbf{29.5\%} of runs whose static DOM already
passes. \bench{} therefore scores a patch against the living system it lands in,
not the target alone.

\end{abstract}

\begin{figure*}[h]
  \centering
  \vspace{-13pt}
  \includegraphics[width=0.8\textwidth]{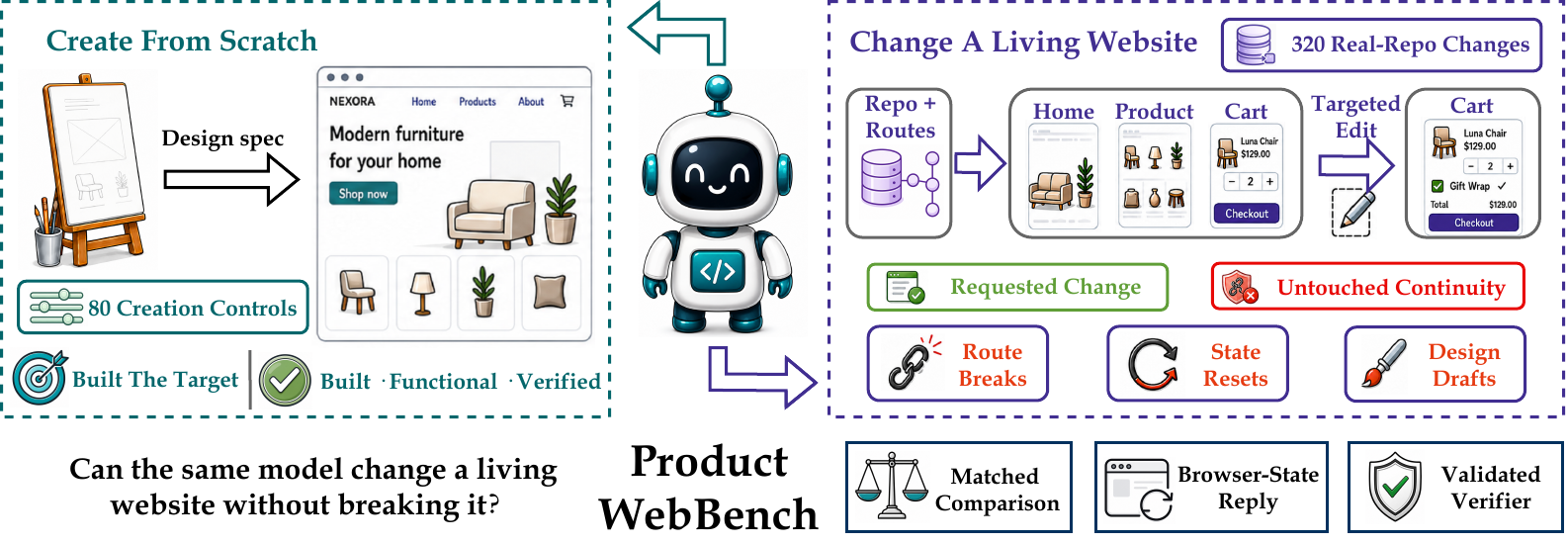}
  \begin{minipage}{0.8\textwidth}
  \vspace{-4pt}
  \caption{Conceptual motivation, not a difficulty-matched comparison. Building
  from scratch asks if the page works; a live change also asks what survives.}
  \vspace{-15pt}
  \label{fig:intro}
  \end{minipage}
\end{figure*}

\section{Introduction}
\label{sec:intro}

Large language models (LLMs) are increasingly capable of modifying existing software repositories from natural-language instructions, enabling agents to assist with tasks such as feature implementation, bug fixing, and repository-level development. Among these tasks, frontend development has a unique challenge because changes to a user interface often propagate beyond the requested functionality. For example, adding a shopping cart control may require modifications to shared states, checkout flows, responsive layouts, and components across different routes. As a result, a successful frontend agent should not only implement the requested feature, but also preserve the existing behaviors and integrity of the website it modifies.

We refer to this requirement as \emph{website continuity}. 
A frontend change can satisfy a local requirement, such as a DOM assertion for the new component, while breaking previously working states, workflows, or visual consistency. 
Therefore, the unit of success is not the patch alone, but the patch together with the website it lands in.
Evaluating such changes requires evidence about the behavior that existed before the modification, not only the final state after the edits (Figure~\ref{fig:intro}).


This is harder still in long-horizon development, where requirements arrive incrementally and accepted changes become the ground later ones are built on. 
A later patch may fix a newly requested feature while silently invalidating an earlier browser state, interaction workflow, or cross-page dependency. 
A continuity evaluation must therefore reason over the evolution of the site: replay established behaviors after subsequent edits, separate newly introduced failures from unresolved existing ones, and hold the acceptance contract stable as requirements evolve.

Existing benchmarks only cover parts of this problem.  
Frontend generation benchmarks evaluate whether agents can produce desired interfaces or behaviors~\citep{si2025design2code,zhang2026miniappbench}, but they do not measure whether modifications preserve an existing application. 
Repository-level software engineering benchmarks evaluate code changes against executable tests~\citep{jimenez2024swebench}, but lack browser-level evidence involving viewports, interaction histories, and shared frontend components. 
Browser-agent benchmarks evaluate agents operating on existing websites~\citep{zhou2024webarena,he2024webvoyager}, but do not study agents that modify the source code defining those websites. 
None of these settings evaluates whether an agent can continuously improve an existing website while preserving previously established behaviors.


To address this gap, we introduce \bench{}, a benchmark for evaluating long-horizon frontend agents under a fixed continuity contract. 
\bench{} contains 320 Change tasks, which evaluate feature modifications and workflow repairs on existing repositories, and 80 Build tasks, which evaluate long-horizon website construction. 
Each task is constructed from a runnable repository together with captured browser evidence, including routes, viewports, actions, and intermediate states. 
During evaluation, a capability-conditioned simulated user reveals requirements progressively and provides feedback while the agent modifies a cumulative workspace. 
This design enables evaluation not only of whether agents complete requested changes, but also whether they preserve the website behaviors established throughout the development process.


We evaluated 8,737 valid trajectories from 13 models under this protocol.
The results reveal a consistent gap between feature completion and website continuity.
Among 674 Change runs that satisfy all requested-work checks, 190 (\textbf{28.2\%}) still fail our continuity checks.
Moreover, \textbf{10.8\%} of these runs regress browser states they had previously passed.
Similarly, among 1,322 runs that successfully satisfy the target DOM requirements, \textbf{29.5\%} fail interaction checks.
These results show that implementing the requested functionality is only one part of successful frontend development; preserving existing website behavior remains a separate challenge.


We summarize our contributions as follows:
\noindent\ding{182}~\textbf{An acceptance contract testable before any solver
runs.} Each of the 400 tasks is admitted only once its reference passes every
gate and authored wrong solutions fail, so satisfiability and gate
discrimination are settled before evaluation, not argued after it.\\
\noindent\ding{183}~\textbf{Long-horizon runs with cumulative gates.} A
simulated user reveals the task one stage at a time, and a cleared stage is
never retired: its assertions stay armed at every later checkpoint, so
continuity is scored across the whole trajectory, not its endpoint.\\
\noindent\ding{184}~\textbf{Two regression channels with exact provenance.} A
state the run was seen to pass, and a gate the untouched repository
satisfied, are two routes to a regression no labelling step has to guess at, and
they are nearly disjoint. Their union catches $\mathbf{18.3\%}$ [$15.3\%$, $21.6\%$] of the
586 covered completing runs damaging a verified state; $13.3\%$ are caught and
still failing a continuity check at the end.


\section{Evidence-Backed Task Construction}
\label{sec:benchmark}

\begin{figure}[!t]
  \centering
  \includegraphics[width=0.8\textwidth]{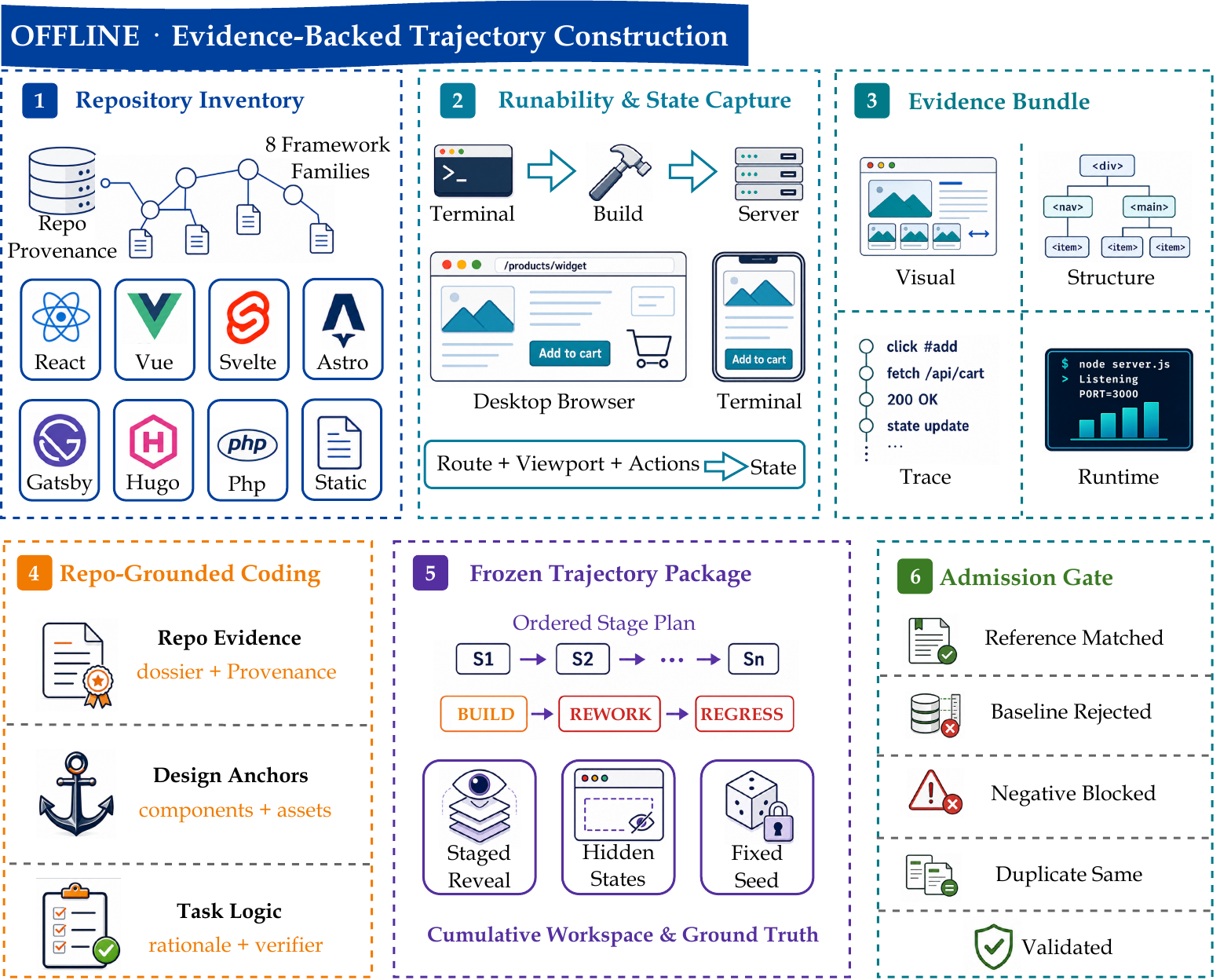}
  \vspace{-5pt}
  \caption{Offline construction. Repository states supply the evidence for ordered
  stages; admission requires a passing reference, a failing baseline and
  smart-wrong control, and a distinct artifact.}
  \vspace{-5pt}
  \label{fig:offline_framework}
\end{figure}

The unit of evaluation in \bench{} is a frozen task package: a repository
snapshot, an ordered change request, and executable evidence of what must hold
after the change. Figure~\ref{fig:offline_framework} shows the construction
pipeline, and Section~\ref{sec:protocol} drives the same package through
evaluation.

\subsection{Task Package and Regimes}

Let a task package be represented as
\begin{equation}
  \mathcal{Q} = (R_0, P, S, E, G, \Delta^{\star}, N),
  \label{eq:task_package}
\end{equation}
where $R_0$ is the repository snapshot with provenance, $P$ the
allowed path scope, $S=(S_1,\ldots,S_K)$ the stage plan, $E$ the
browser evidence and $G$ the deterministic checks. $\Delta^{\star}$ is
the reference patch and $N$ the negative controls; both are used for
admission only and never reach the solver.

\textbf{Task regimes and coverage.} The locked suite contains 400 tasks in two
regimes. The 320 \textbf{Change} tasks begin from a runnable repository and ask
the agent to modify it while preserving incumbent behavior; the 80
\textbf{Build} tasks grow a product site through 5 to 14 ordered milestones, a
construction and long-horizon regime rather than an easy control for Change.
Each Change task is organized around one \emph{facet} -- the place where
continuity can break: state, design, experience, content and assets, regression
or implementation -- and the facet fixes which evidence has to survive the
change. Every candidate repository is
installed, built, served and exercised before authoring, and what cannot be
reproduced or legally packaged never enters the pool; the inventory spans React,
Vue, Svelte, Astro, Gatsby, Hugo, PHP and static sites.
Appendix~\ref{app:inventory} gives the capability labels and the full facet
inventory, and Appendix~\ref{app:results} the finer framework breakdown.

\subsection{Browser Evidence and Stage Construction}

Authors capture $R_0$ before
writing the request. Each browser state in $E$ is identified by a route,
viewport and action prefix and carries screenshots, visual anchors, DOM
structure and accessible names; retaining the action prefix makes a state
replayable, since a cart after adding an item is a different target from the
same route before that action.

The request is then grounded in $E$: design requirements point to repository-native assets and tokens, behavior
requirements name a route, viewport, action prefix and expected state, and
preservation requirements identify untouched states sharing navigation, data,
styling or source files with the target. Those states compile into regression
gates replayed at every checkpoint, and the plan $S$ follows the dependency
structure in the evidence, so a cart task exposes route access, item state, a new
control, cross-page persistence, responsive behavior and regression replay, one
per turn.

\subsection{Frozen Contract and Admission Controls}

Once authored, the fields of $\mathcal{Q}$ are frozen together, so changing the
solver or the conversation that presents a task never changes its acceptance
checks or the evidence behind them. Admission then tests whether the package
discriminates the intended behavior: applying $\Delta^{\star}$ to $R_0$ must
pass every applicable check, and plausible wrong implementations in $N$ must
fail: a visually convincing control with inert state, a responsive layout that
overflows at one viewport, a local edit that drops an existing route. One more
property is observed, not asserted: the task must not already be done. Of the 234
slots whose untouched repository renders under the frozen toolchain, the
repository handed to the solver passes every gate on \textbf{0} of them, the
closest reaching $17/19$. On every one of them, then, the requested work is
measurably outstanding when the solver starts. Section~\ref{sec:experiments} reports how the authoring
pool scores on both halves of this test, and Appendix~\ref{app:verifier} the
audit behind them.

\section{Cumulative Multi-Turn Evaluation}
\label{sec:protocol}

\begin{figure}[!b]
  \centering
  \vspace{-8pt}
  \includegraphics[width=0.8\textwidth]{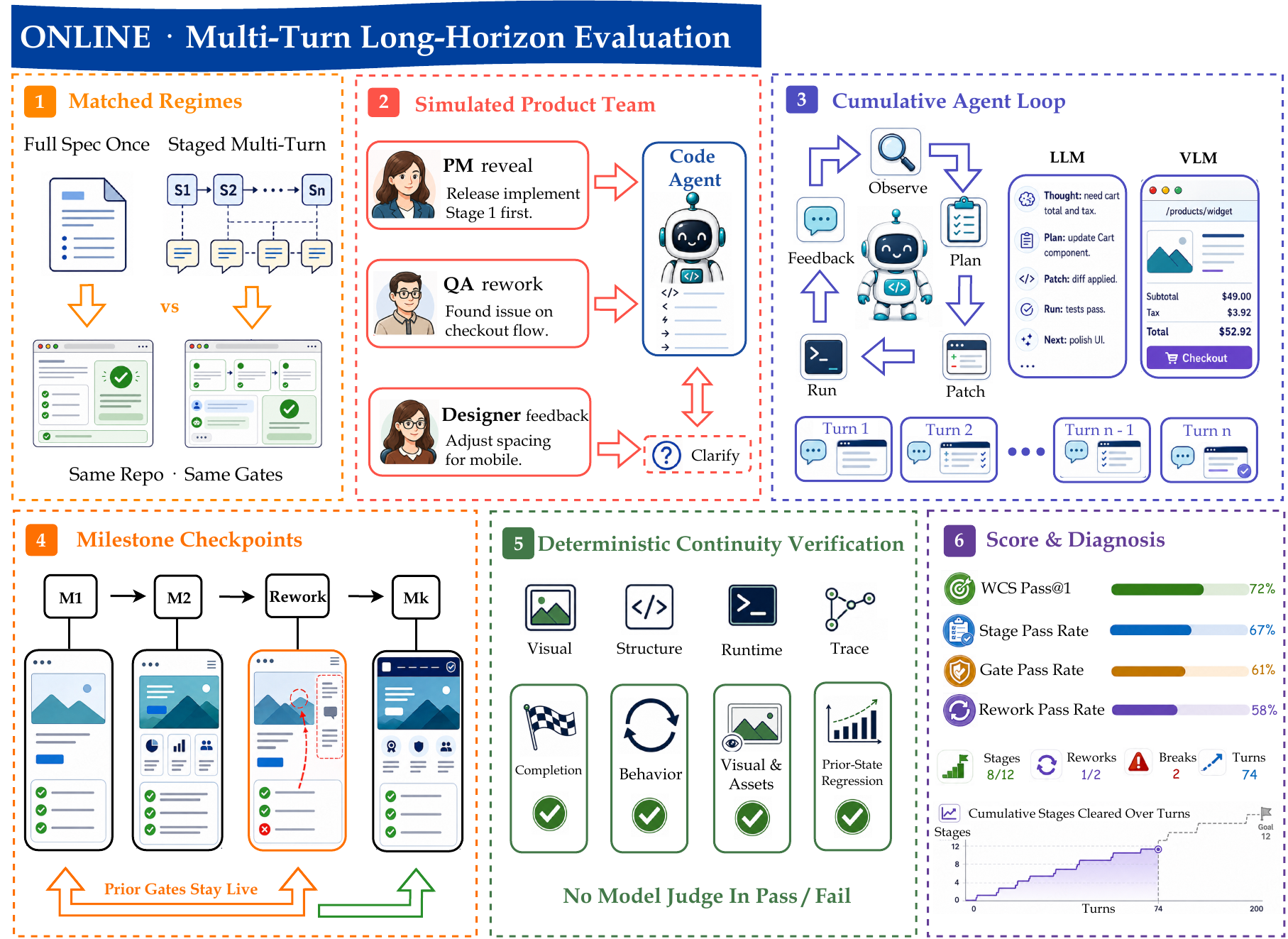}
  \vspace{-5pt}
  \caption{Online multi-turn evaluation. PM/QA/designer are alternative personas,
  not concurrent. The harness checks each milestone and replays every prior gate;
  inset values are schematic.}
  \label{fig:online_framework}
\end{figure}

Evaluation starts from $R_0$ and walks the stage plan $S$ of
Eq.~\ref{eq:task_package}. One simulated user presents the work, the solver
edits a cumulative workspace, and the harness checks the result against the
frozen checks $G$ (Figure~\ref{fig:online_framework}). One design choice does
most of the work: clearing a milestone does not retire it. Every assertion a
stage clears becomes a standing obligation retested after every later edit, so
the score reflects the whole trajectory and not just the state reached at the final turn.

\subsection{Task Presentation and Simulated User}

Two interfaces read the same
frozen package: the \textbf{staged} regime reveals it as a sequence of work
units, the \textbf{full-spec} regime hands the agent everything at once. Every
model result is staged, while the agent-scaffold study uses full-spec because
CLI agents own their turn loop, so the two arms are reported separately and
never compared. In the staged regime each task selects a persona
from its capability (PM for Build and Edit, QA for Repair, designer for
Consistency), and the persona controls only the wording that reveals work
units authored in advance. It does not generate requirements, read the reference
patch, edit the verifier or decide pass/fail, and reveal, solver, checkpoint
and pushback events are logged separately (Appendix~\ref{app:interfaces}).

\subsection{Workspace Updates and Milestone Replay}

Each turn the solver
receives scoped source context and returns a message plus structured edits. The
harness applies accepted edits to the cumulative workspace, builds and serves
the project, captures browser states and runs the deterministic checks,
returning grounded failure text and, on the vision leg, the latest screenshot;
the driver allows 12 turns per stage and 200 per run. A milestone clears only
when the current workspace builds, serves, reaches the specified route, viewport
and action state, and passes its assertions, and once cleared its gates join the
active prior-state set. Let $G_j\subseteq G$ denote the checks attached to stage
$S_j$, and let $C_t$ contain the indices of stages cleared before checkpoint
$t$. The checks retained for replay are
\begin{equation}
  G_t^{\mathrm{prior}} = \bigcup_{j\in C_t}G_j .
  \label{eq:prior_gates}
\end{equation}
Every later checkpoint replays these alongside the current milestone's
requirements, so a patch that fixes the current stage but invalidates an earlier
state is recorded as a regression. Rework reuses the workspace: later stages
inherit the successful edits and the unresolved state alike.

\begin{table*}[!t]
\centering
\scriptsize
\setlength{\tabcolsep}{2.0pt}
\renewcommand{\arraystretch}{0.99}
\newcommand{\resultgroup}[1]{%
  \arrayrulecolor{black!65}\specialrule{0.055em}{0pt}{0pt}%
  \rowcolor{HeaderColor}
  \multicolumn{13}{c}{\rule[-0.30ex]{0pt}{2.05ex}\smash{\textbf{#1}}}\\[-0.8pt]
  \arrayrulecolor{black!65}\specialrule{0.055em}{0pt}{0pt}%
  \arrayrulecolor{black}%
}
\caption{Main results on \bench. \textbf{Task\%} is the strict whole-task rate
of Section~\ref{sec:protocol}; capability columns use labelled cells only
(Appendix~\ref{app:inventory}). Best per column \textbf{bold}.}
\label{tab:main_pwb_modb}\resizebox{\textwidth}{!}{%
\begin{tabular}{lrrrrrrrrrrrr}
\toprule
\multirow{2}{*}{\textbf{Model}} &
\multicolumn{3}{c}{\textbf{Task\%}} &
\multicolumn{1}{c}{\textbf{Gate\%}} & \multicolumn{1}{c}{\textbf{Stage\%}} &
\multicolumn{3}{c}{\textbf{Coverage \& \#Pass}} &
\multicolumn{4}{c}{\textbf{Gate\% by capability}} \\
\cmidrule(lr){2-4} \cmidrule(lr){5-5} \cmidrule(lr){6-6} \cmidrule(lr){7-9} \cmidrule(lr){10-13}
& \textbf{Avg} & \textbf{L} & \textbf{V}
& \textbf{Overall} & \textbf{Overall}
& $n_L$ & $n_V$ & \textbf{\#Pass}
& \textbf{Edit} & \textbf{Repair} & \textbf{Consist.} & \textbf{Build} \\
\resultgroup{Closed-source models}
\rowcolor{ClosedColor}
Claude-Opus-4.8 & \textbf{3.50} & \textbf{3.75} & \textbf{3.25} & 59.49 & 41.69 & 400 & 400 & \textbf{28}/800 & 58.70 & 64.72 & 55.08 & 62.43 \\
\rowcolor{ClosedColor}
GPT-5.5 & 2.75 & 3.00 & 2.50 & \textbf{61.71} & \textbf{42.45} & 400 & 400 & 22/800 & \textbf{59.49} & 64.90 & \textbf{57.64} & \textbf{68.52} \\
\rowcolor{ClosedColor}
Claude-Sonnet-4.6 & 1.62 & 1.25 & 2.00 & 49.41 & 32.65 & 400 & 400 & 13/800 & 50.00 & 55.62 & 47.71 & 43.49 \\
\rowcolor{ClosedColor}
Gemini-3.5-Flash & 1.25 & 1.25 & 1.25 & 52.43 & 34.88 & 400 & 400 & 10/800 & 52.88 & 60.31 & 48.59 & 49.41 \\
\resultgroup{Open-source models}
\rowcolor{GoodColor}
DeepSeek-V4-Flash & 3.25 & 3.25 & --- & 56.64 & 38.24 & 400 & --- & 13/400 & 56.88 & 63.44 & 51.50 & 58.09 \\
\rowcolor{GoodColor}
Kimi-K3 & 2.93 & 2.75 & 3.10 & 54.37 & 39.30 & 400 & 387 & 23/787 & 53.72 & 63.45 & 50.38 & 52.76 \\
\rowcolor{GoodColor}
GLM-5.2-FP8 & 2.53 & 2.53 & --- & 53.78 & 39.90 & 395 & --- & 10/395 & 53.76 & \textbf{65.61} & 50.09 & 47.70 \\
\rowcolor{GoodColor}
Hunyuan-3-Preview & 2.34 & 2.34 & 2.34 & 50.03 & 34.69 & 385 & 385 & 18/770 & 52.15 & 57.13 & 47.72 & 43.52 \\
\rowcolor{GoodColor}
Qwen3.7-Max & 2.25 & 2.25 & --- & 54.82 & 37.32 & 400 & --- & 9/400 & 54.25 & 62.73 & 47.82 & 57.70 \\
\rowcolor{GoodColor}
Qwen3.6-35B-A3B & 2.03 & 1.80 & 2.25 & 50.05 & 33.46 & 388 & 400 & 16/788 & 50.95 & 59.61 & 46.03 & 45.49 \\
\rowcolor{GoodColor}
Kimi-K2.7-Code & 2.01 & 2.26 & 1.75 & 52.99 & 38.10 & 398 & 399 & 16/797 & 53.04 & 59.32 & 48.42 & 53.92 \\
\rowcolor{GoodColor}
DeepSeek-V4-Pro-Preview & 2.00 & 2.00 & --- & 54.34 & 37.82 & 400 & --- & 8/400 & 53.59 & 58.32 & 51.50 & 56.58 \\
\rowcolor{GoodColor}
MiniMax-M3 & 1.50 & 1.75 & 1.25 & 54.42 & 36.42 & 400 & 400 & 12/800 & 56.10 & 62.52 & 48.91 & 52.96 \\
\bottomrule
\end{tabular}}
\vspace{-15pt}
\end{table*}

\subsection{Verification and Continuity Score}

The final verifier runs completion,
behavior, visual/asset, runtime and prior-state checks, and
Appendix~\ref{app:gates} lists all 19 signals and the three families that
partition them. Whole-task success is intentionally strict, so we also report
the graded \textbf{Website Continuity Score} (WCS). For attempt $i$, let $s_i$
be the fraction of \emph{stages} cleared, counting stages only and not reworks;
$g_i$ the deterministic gate pass rate; $v_i$ the binary final gate pass; and
$b_i$ the number of regression breaks, counted as one per (checkpoint, gate)
pair recorded as regressed:
\begin{equation}
\wcs_i = \left[0.5s_i + 0.3g_i + 0.2v_i - 0.05b_i\right]_+ .
\label{eq:wcs}
\end{equation}
Here $[x]_+=\max(0,x)$, so WCS retains partial progress while penalizing
regressions. The weights and the per-break penalty are design choices, and
Appendix~\ref{app:results} gives a worked example and what the per-checkpoint
accumulation of $b_i$ does to the graded scale. Every \emph{rate} we report on
the regression channels uses only the indicator $b_i>0$ and is unaffected by
that accumulation.

Three success notions recur and are not interchangeable. \textbf{Requested-work
completion} means every completion-family gate passes. \textbf{Final gate pass}
means all nineteen gates pass on the final workspace, which 259 of the 8,737
cells reach; it is the $v_i$ of Eq.~\ref{eq:wcs} and what the full-spec
scaffold arm reports, since that arm verifies only once. \textbf{Whole-task pass} additionally
requires every stage and rework to clear with $b_i=0$. It is the stricter
notion, it holds on 198 cells, and it is the \textbf{Task\%} column of
Table~\ref{tab:main_pwb_modb}. Case~(1) in
Appendix~\ref{app:case_studies} passes all nineteen final gates on its delivered
workspace without clearing every unit, which is exactly the gap between the
second notion and the third.

The gates are also partitioned by what they measure. The \textbf{completion}
family asks whether the requested work is present. The rest are the
\textbf{continuity checks} this paper reports throughout: regression text,
visual anchors, layout, state artifacts and runtime health, read conservatively
by excluding the general \texttt{visual\_regression} signal. An attempt is
\textbf{complete-but-not-cleared} when it passes every completion gate and fails
at least one continuity check (Appendix~\ref{app:gates}).

\section{Experiments}
\label{sec:experiments}

\begin{table*}[!t]
\centering
\small
\setlength{\tabcolsep}{3pt}
\renewcommand{\arraystretch}{1.16}
\caption{Agent scaffolds on \bench, grouped by backbone. Each row runs 400
tasks; capability columns report Gate\%. Stage\% needs checkpoints, undefined
here. Best overall in \textbf{bold}.}
\label{tab:cli_scaffold}
\setlength{\tabcolsep}{9pt}
\begin{tabular}{lrrrrrr}
\toprule
\textbf{Agent} & \textbf{Task\%} & \textbf{Gate\%} &
\textbf{Edit} & \textbf{Repair} & \textbf{Consist.} & \textbf{Build} \\
\midrule
\rowcolor{HeaderColor}[0pt][\tabcolsep]
\multicolumn{7}{@{}l}{\textit{Backbone: Claude Sonnet 4.6}} \\
\addlinespace[2pt]
\rowcolor{SonnetTint}
mini-swe-agent & \textbf{0.75} & 64.58 & 64.24 & 69.13 & 61.16 & 67.63 \\
\rowcolor{SonnetTint}
Claude Code & 0.50 & 62.51 & 62.13 & 67.07 & 57.80 & 66.84 \\
\rowcolor{SonnetTint}
Qwen Code & 0.50 & 59.80 & 58.41 & 65.65 & 56.40 & 61.84 \\
\rowcolor{SonnetTint}
Goose & 0.50 & 61.93 & 62.30 & 66.86 & 58.98 & 62.63 \\
\rowcolor{SonnetTint}
Kimi CLI & 0.25 & 61.28 & 61.33 & 65.22 & 60.25 & 60.26 \\
\rowcolor{SonnetTint}
Aider & 0.00 & 57.64 & 56.01 & 61.38 & 51.68 & 66.45 \\
\rowcolor{SonnetTint}
opencode & 0.00 & 52.03 & 51.49 & 56.26 & 48.05 & 56.78 \\
\midrule
\rowcolor{HeaderColor}[0pt][\tabcolsep]
\multicolumn{7}{@{}l}{\textit{Backbone: GPT-5.5}} \\
\addlinespace[2pt]
\rowcolor{GPTTint}
mini-swe-agent & 0.50 & \textbf{69.68} & \textbf{67.51} & 70.77 & \textbf{64.93} & 80.66 \\
\rowcolor{GPTTint}
Codex & 0.50 & 68.25 & 67.33 & 71.41 & 63.52 & 74.87 \\
\rowcolor{GPTTint}
Goose & 0.50 & 69.22 & 65.22 & \textbf{71.62} & 63.84 & \textbf{81.78} \\
\rowcolor{GPTTint}
Aider & 0.25 & 60.75 & 58.70 & 63.73 & 53.36 & 73.82 \\
\rowcolor{GPTTint}
opencode & 0.00 & 52.80 & 51.83 & 57.04 & 48.64 & 58.16 \\
\bottomrule
\end{tabular}
\vspace{-8pt}
\end{table*}

\begin{table}[!htb]
\centering
\sbox{\pwbtab}{\small
\setlength{\tabcolsep}{3pt}%
\renewcommand{\arraystretch}{1.10}%
\begin{tabular}{lrr}
\toprule
\rowcolor{HeaderColor}
& \multicolumn{2}{c}{\textbf{Requested work}} \\
\rowcolor{HeaderColor}
\textbf{Additional checks} & \textbf{Complete} & \textbf{Incomplete} \\
\midrule
Pass & 484 & 1{,}154 \\
Fail & \textbf{190} & 5{,}151 \\
\bottomrule
\end{tabular}}%
\begin{minipage}[c]{\dimexpr\textwidth-\wd\pwbtab-1.2em\relax}
\raggedright
The quadrant decomposition shows what target-only scoring omits: a model can
produce the requested state while failing an additional route, responsive,
anchor, state or runtime check. State-level provenance is needed to say which of
these are confirmed incumbent-state regressions; completion alone credits all 190
attempts that fail the conservative checks.
\end{minipage}\hfill
\begin{minipage}[c]{\wd\pwbtab}
\small
\setlength{\abovecaptionskip}{0pt}%
\setlength{\belowcaptionskip}{4pt}%
\caption{Completion versus additional checks, 6,979 Change cells.}
\label{tab:completion_continuity}
\usebox{\pwbtab}
\end{minipage}
\end{table}

\subsection{Setup}

We evaluate 13 frontier and open coding models under the staged protocol of
Section~\ref{sec:protocol}: every run starts from a pristine repository with the
same stage plan, gates, allowed paths and budgets, cannot reach the reference or the
negative controls, and is checked by the frozen 19-gate verifier at every
checkpoint. Nine of the 13 also have image-valid screenshot runs paired with
their own text-only cells. A verdict counts
only if it postdates its task's frozen specification, leaving \textbf{8,737 live
cells}, 6,979 Change and 1,758 Build, the single snapshot every number in this
paper is computed from. Before interpreting any model failure we check that the
benchmark can separate a correct solution from a plausible regression: across
the validated authoring pool all 479 reference implementations pass and all 479
smart-wrong controls fail, Wilson [0.992, 1.000] in both directions. That is an
admission property of the pool, not an accuracy estimate on unseen solver
output. What it rules out is that a low score comes from an impossible
reference, or a gate that lets the authored bad state through: the contract
discriminates before any model runs. Appendix~\ref{app:results} gives the
models, per-question denominators and statistical conventions, and
Appendix~\ref{app:verifier} measures accuracy on solver output.

\subsection{Main Results}

\textbf{Whole-task passes are sparse; the graded scale is not.} Only 198 of the
8,737 cells clear every stage and rework, pass every gate, and record no
regression break, and whole-task rates span $1.25$ to $3.50\%$, too narrow a
band to order the field. Gate\% spans $49.41$ to $61.71$ over the same runs and does order it,
and the two disagree at the top: Claude-Opus-4.8 leads whole-task success while
GPT-5.5 leads Gate\% and Stage\%. By capability the ordering is the same for all
13 models, Repair highest and Consistency lowest of the three Change
capabilities, which asks a change to match conventions that live in the
repository rather than in the request.

\textbf{Scaffolds change the number as much as backbones do.}
Table~\ref{tab:cli_scaffold} runs eight distinct scaffolds over the same 400
tasks, four of them under both backbones, so scaffold and backbone can be
compared as one statistic rather than a range against a mean. On that matched
$4\times2$ grid, swapping the scaffold under a fixed backbone moves the
component mean by $8.06$~pt on Sonnet-4.6 and $10.52$~pt on GPT-5.5, while
swapping the backbone under a fixed scaffold moves it by $5.08$~pt. How an agent
is wired around a model therefore matters as much as which model it is, which is
a reason to report the scaffold with the score
(Appendix~\ref{app:results} defines the component mean and states what this
matched grid does and does not control).


\section{Analysis}
\label{sec:results}

\begin{figure}[t]
\centering
\includegraphics[width=0.99\linewidth]{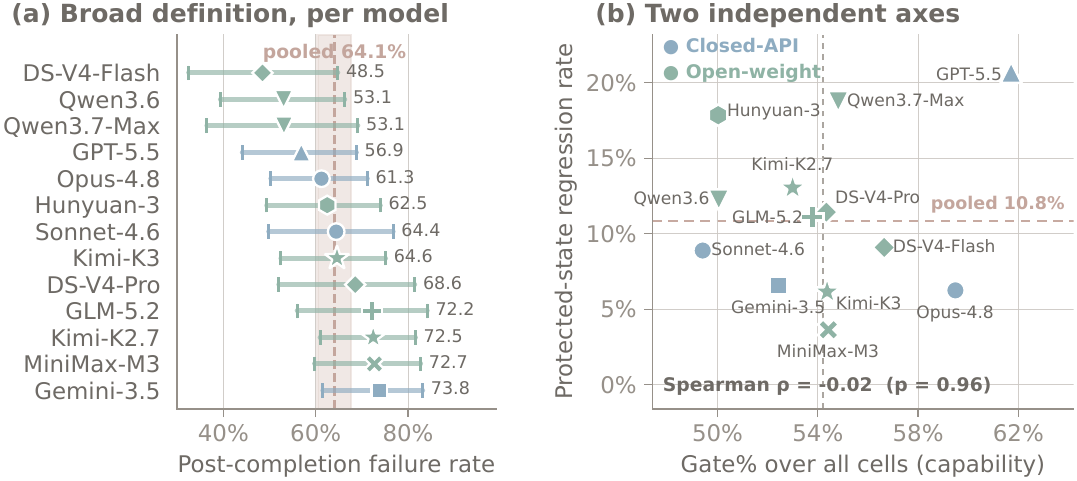}
\vspace{-6pt}
\caption{Thirteen models, 674 completing Change runs. (a) Post-completion
failure, Wilson-95. (b) Regression rate against Gate\%: unrelated orderings.}
\vspace{-10pt}
\label{fig:done_but_broke}
\end{figure}

\sbox{\pwbtab}{\small
\setlength{\tabcolsep}{5pt}%
\renewcommand{\arraystretch}{1.12}%
\begin{tabular}{lrr}
\toprule
\rowcolor{HeaderColor}
\textbf{Observation point} & \textbf{Count} & \textbf{Rate} \\
\midrule
Replay, any checkpoint & 73/674 & \textbf{10.8\%} \\
\midrule
Class (a): initial state & 66/586 & 11.3\% \\
Class (b): run's own state & 57/586 & 9.7\% \\
\quad both channels & 16/586 & 2.7\% \\
Either channel & 107/586 & \textbf{18.3\%} \\
\quad still failing at the end & 78/586 & \textbf{13.3\%} \\
\bottomrule
\end{tabular}}%
\begin{wraptable}[9]{r}{\wd\pwbtab}
\vspace{-\intextsep}\vspace{-16pt}
\small
\setlength{\abovecaptionskip}{0pt}%
\setlength{\belowcaptionskip}{4pt}%
\caption{Provenance-exact regression. Full rows and intervals:
Table~\ref{tab:regression_endpoints_full}.}
\label{tab:regression_endpoints}
\usebox{\pwbtab}
\end{wraptable}
\textbf{Completion is not continuity.}
Table~\ref{tab:completion_continuity} separates the two verdicts: of the Change
runs that pass every completion-family gate, $\mathbf{28.2\%}$ still fail at
least one continuity check, rising to 64.1\% under the broad visual-inclusive
reading. Every model shows it (Figure~\ref{fig:done_but_broke};
Appendix~\ref{app:regression} gives the intervals and the per-model spread).
Nor are these runs failing for
want of a patch that applies: 98.3\% of failing cells land at least one edit.
And an element being present does not make it work: on the interactive cells
whose static DOM requirement already passes, interaction assertions still fail
29.5\% of the time
(Table~\ref{tab:diagnostics}; Appendix~\ref{app:results}).

\begin{figure}[!tb]
\centering
\includegraphics[width=0.99\linewidth]{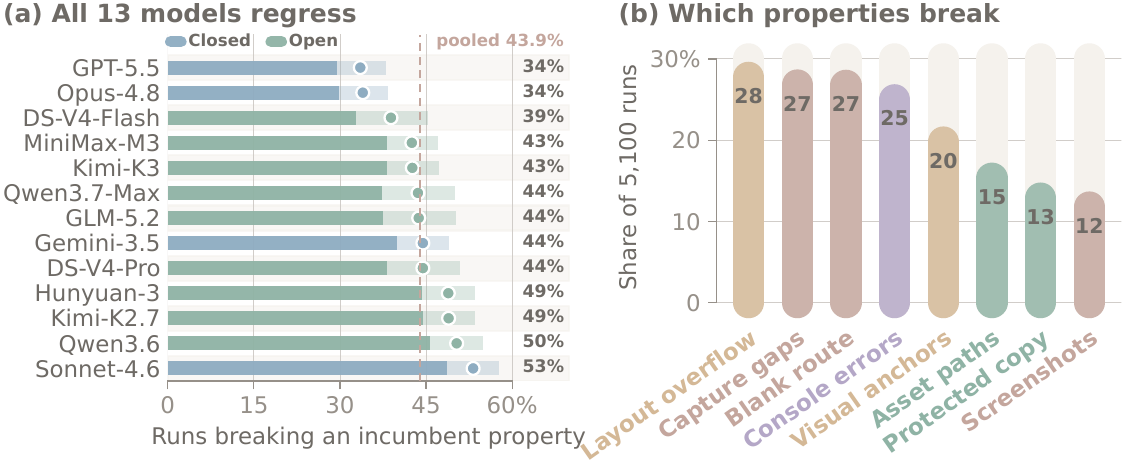}
\vspace{-5pt}
\caption{The untouched-repository channel. (a) Per-model rate: dot, Wilson-95
band, pooled 43.9\% dashed. (b) What breaks, of 5,100 covered runs; colour marks
the family.}
\vspace{-4pt}
\label{fig:pristine_regression}
\end{figure}

\textbf{Two regression channels with exact provenance.} One channel needs no
target-versus-prior labelling at all: whenever a stage clears, its assertions
enter a protected set the harness replays at \emph{every} later checkpoint of
the same run (Section~\ref{sec:protocol}), so an assertion that passes at stage
$k$ and fails at $k'>k$ is a pass-to-fail transition on a state the run itself
established. Among completing runs this fires on $\mathbf{10.8\%}$.
That set can only hold states a run established itself, so for the complementary
class we captured each repository as handed to the solver and ran the
same frozen suite on it: a gate the untouched repository passes and the final
workspace fails is a regression with exact provenance whether or not the run
asserted it. It fires on 43.9\% of covered Change runs, no model escapes, and the rate falls
as model quality rises (Spearman $\rho=-0.88$ against mean WCS). What it catches
is ordinary product health: horizontal overflow, a route that stops rendering,
console errors. On the runs where both channels can be read they are close in
size and nearly disjoint, so together \textbf{18.3\%} are caught damaging a
state that was verified working and $\mathbf{13.3\%}$ are both caught and still
failing a conservative check at the end: the first speaks to trajectories, the
second to delivered sites. Neither restates the post-completion gap
(Figure~\ref{fig:pristine_regression} and Table~\ref{tab:regression_endpoints};
Appendix~\ref{app:regression} gives the coverage, the residual cases and the
gates each channel rests on, which bounds the claim).

\begin{figure}[!tb]
\centering
\includegraphics[width=0.99\linewidth]{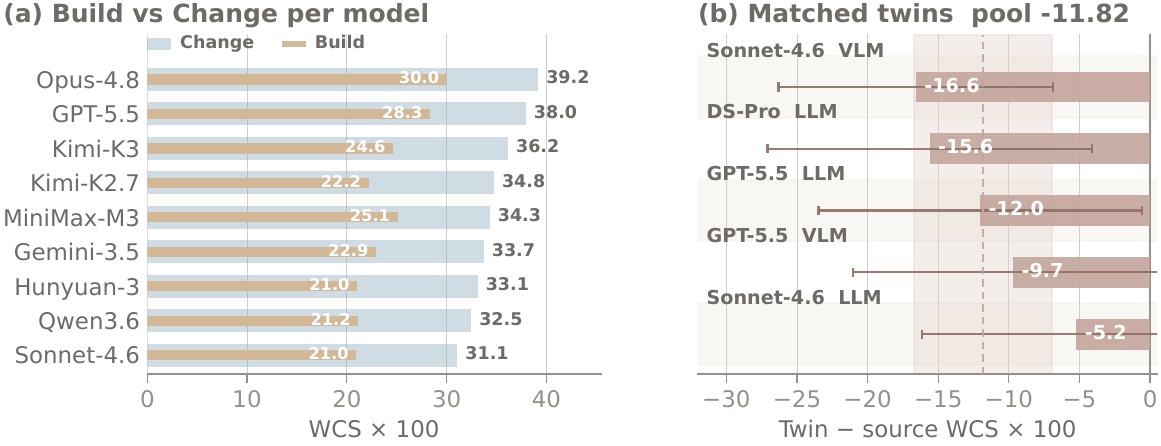}
\vspace{-7pt}
\caption{(a) Build versus Change raw WCS per model; the Build bar is shorter
for all nine. (b) Paired twin-minus-source change per model-leg, pooled dashed;
stage count matched by construction.}
\vspace{-20pt}
\label{fig:regime_and_twins}
\end{figure}

\textbf{Long-horizon construction.} Build and Change
occupy different regimes: on the nine-model dual-modal panel pooled Build WCS is
0.240 against 0.348 for Change, the gap sits in stage completion, not gate
quality, and Build plans are the longer ones (Table~\ref{tab:regimes}). Longer
runs are not recovering ones: on Build tasks of at least seven stages the
correlation between turns and stages cleared is negative in every stage-count
bucket, median $-0.91$, because cleared units finish fast while a failed one
burns all 12 turns (Figure~\ref{fig:turn_thrash};
Appendix~\ref{app:limitations} covers the stopping rule).
Creation is also harder than the matched in-place change. Because Build
plans are longer, the pooled gap alone cannot separate regime from trajectory
length, so we settle it with an authored twin study: twenty-one Change tasks
duplicated as \emph{greenfield twins} that reach the same end state from an
empty baseline, on the same site, under identical acceptance gates and with
stage counts matched by construction. Across the 105 paired cells the pooled
twin$-$source $\Delta\mathrm{WCS}$ is $-11.82$ points on the $100\times$WCS
scale ($t=-4.74$), with all five model-legs negative. The two
regimes are related but not interchangeable, and the capability slices do not
order the field identically either: ranking the 13 models by Gate\% within each
capability, GPT-5.5 leads Edit, Consistency and Build while GLM-5.2-FP8 leads
Repair
(Figures~\ref{fig:regime_and_twins} and~\ref{fig:build_reshuffles};
Appendix~\ref{app:results} reports the rank coherence across the four slices,
which is high but not uniform).

\begin{figure}[!tb]
\centering
\includegraphics[width=0.95\linewidth]{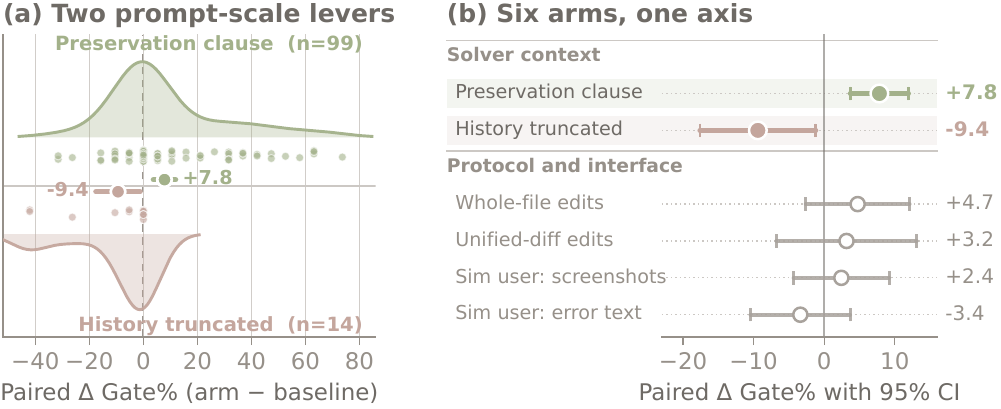}
\vspace{-6pt}
\caption{Paired change in Gate\% per arm. (a) The two prompt-scale levers,
mirrored, one dot per slot. (b) The six arms with 95\% CIs; only what the
solver is told clears zero.}
\vspace{-5pt}
\label{fig:history_preservation}
\end{figure}

\textbf{What moves the continuity signal.} Capability does not
track protected-state regression here: ranking the 13 models by Gate\% and by
the share of their completing runs that break a protected state gives two
orderings with no detectable relation at this sample size (Spearman
$\rho=-0.02$, $p=0.96$; Figure~\ref{fig:done_but_broke}b), and the strongest
model on Gate\% carries the highest regression rate of all. What does
move the graded gate rate is prompt scale, in both directions. Truncating the solver's
history to five turns drops $\text{Gate\%}$ by
$9.40$ points while stage completion and turn count are essentially unchanged,
so the solver reaches the same stages and does worse work at them. Adding a ``do not break unrelated behavior; keep everything else
byte-for-byte'' clause lifts $\text{Gate\%}$ by $7.81$~pt
without costing a whole-task pass
(Figure~\ref{fig:history_preservation}). The simulated user is not doing the
solver's work: changing what it says, in either direction, moves nothing
distinguishable from zero, and neither does the edit format nor giving the
solver screenshots. What the solver is told moves the score; what it is shown,
and how the request is worded, do not. Appendix~\ref{app:results} tabulates
these arms with the solver-vision contrast
Figure~\ref{fig:history_preservation} leaves out, and
Appendix~\ref{app:regression} the endpoint these arms cannot resolve.

\section{Related Work}
\label{sec:related}

\textbf{Frontend and repository benchmarks.} Benchmarks connect webpage
designs with code generation and visual reconstruction
\citep{si2025design2code,yun2024web2code,gui2025webcode2m}, then expand to
framework-based editing and repair \citep{xiao2025designbench} and to
the functional, visual and interactive quality of generated applications
\citep{zhang2026miniappbench,lu2025webgenbench,zhang2025artifactsbench}. Closest
in coverage, WebCompass unifies text-, image- and video-conditioned
generation, editing and repair, judging edits by model and generation by browser
agent \citep{lei2026webcompass}. SWE-bench evaluates issue resolution with
executable tests, regression tests included
\citep{jimenez2024swebench}, and SWE-bench Multimodal extends it to visual,
user-facing JavaScript software \citep{yang2025swebenchmultimodal}; agent
interfaces and execution platforms support editing and adaptation to
observations \citep{yang2024sweagent,wang2024codeact,wang2025openhands}, and
complementary work studies refinement, reflection and the limits of self-repair
\citep{madaan2023selfrefine,shinn2023reflexion,olausson2024selfrepair}. Lifecycle
coverage, executable feedback and preservation checks are shared concerns.

\textbf{Browser environments, simulated users, and what \bench{} adds.} Web-agent
research contributes demonstrations, executable environments and grounded
tasks
\citep{deng2023mind2web,zhou2024webarena,koh2024visualwebarena,he2024webvoyager},
OSWorld covers desktop and cross-application workflows
\citep{xie2024osworld}, ReAct studies interleaved reasoning and action
\citep{yao2023react}, AgentBoard measures intermediate progress
\citep{ma2024agentboard}, and simulated-user benchmarks examine tool use against
verifiable outcomes \citep{yao2024taubench,barres2025tau2bench}. What \bench{}
adds is cumulative verification under a frozen contract. A task package binds
each requirement to route, viewport and action-history evidence; a simulated user
stages the work while frozen checks decide acceptance; and every cleared state is
replayed at every later checkpoint, so an edit is graded against the site it
changes, not the target alone. One trajectory therefore yields requested
completion, additional verification failure and provenance-exact regression as
three separate quantities, so progress, persistent failure and damage are told
apart, not inferred from repair turns.


\section{Conclusion}
\label{sec:conclusion}

\bench{} evaluates frontend patches against a living system, not an isolated
target. Its 400 evidence-backed tasks combine real repositories, staged
persona-conditioned interaction, cumulative browser checkpoints and validated
verifiers. Across 13 models, 28.2\% of the Change runs that pass every completion gate
still fail an additional conservative check. Protected-state replay certifies
10.8\% of them as pass-to-fail regressions on states the run established itself,
and the untouched-repository channel raises provenance-exact damage to 18.3\% on
the 586 runs where both are observable, with 13.3\% of that same 586 still
failing a conservative check at the end. Scaffolds matter as much as backbones: on the
full-spec arm, swapping the scaffold under a fixed backbone moves the score
further than the reverse, so the harness belongs in any report of agent
capability. Target completion alone is therefore
insufficient evidence that the site the change lands in still holds.
Appendix~\ref{app:discussion} draws out what these numbers imply for reporting,
Appendix~\ref{app:limitations} bounds the claims, and
Appendix~\ref{app:case_studies} walks seven released trajectories end to end,
one per capability.

\section*{Ethics Statement}
This work evaluates code-generating agents on software repositories and involves
no human subjects, no personal data, and no user studies. Every task is built
from a publicly available frontend repository, and each released package retains
the upstream repository's provenance and license; we redistribute task metadata,
browser-state evidence, and verifier definitions rather than relicensed source.
Evaluation runs in a sandboxed workspace with no deployment to public services
and no production credentials, and provider keys, billing metadata, and
transient workspaces are excluded from the release
(Appendix~\ref{app:reproducibility}). The benchmark measures whether a patch
preserves behaviour that a repository already had; it is a software-quality
instrument and we are not aware of a dual-use risk beyond that of the public
repositories it is built from. The scores we report describe systems under one
fixed protocol and should not be read as general capability claims about any
vendor's model, and we state throughout where provider availability makes model
coverage unbalanced.

\section*{Reproducibility Statement}
Every number in the paper is computed from one frozen result manifest, and that
manifest is what we release: one row per evaluated cell, carrying the stage and
gate components of the score, the applied and rejected edits, the failing gates,
and the normalized capability, facet and framework labels. Rows superseded by a
later submission specification are already removed from it, so the released
manifest is exactly the 8,737 cells every number is taken over.
Section~\ref{sec:experiments} fixes the protocol, budgets, and
stale-row filter; Appendix~\ref{app:gates} gives the gate families,
Appendix~\ref{app:interfaces} the two evaluation arms and their scoring
conventions, Appendix~\ref{app:verifier} the verifier-validation procedure, and
Appendix~\ref{app:regression} the exact definition of both continuity
statistics. Appendix~\ref{app:reproducibility} lists what each released task
package contains and the safeguards applied before release. Rerun variability is
characterised in Appendix~\ref{app:verifier}: the graded metric carries a
per-cell noise floor of $\sigma\approx0.09$, which we quantify and attach to
every graded number, so exact per-cell reproduction of a graded score should not
be expected even at $T=0$; the binary verdict was stable in all fifteen replayed
pairs.

\section*{Use of Large Language Models}

We used a general-purpose LLM assistant for writing assistance only: copy-editing
prose we had already drafted and checking LaTeX and figure formatting. It had no
part in the research itself, and the authors have verified the full text and take
responsibility for its contents.

\bibliography{iclr2027_conference}
\bibliographystyle{iclr2027_conference}

\clearpage
\appendix
\section{Task and Facet Inventory}
\label{app:inventory}

Table~\ref{tab:facet_inventory} reports primary continuity-facet metadata for
the locked Change split. Three tasks have no normalized facet; they are
excluded from facet-stratified analyses and kept in aggregate totals.

\begin{table}[h]
\centering
\begin{minipage}[t]{0.50\textwidth}
\vspace{0pt}
\raggedright
Capability labels come from the frozen task record, so a per-capability column
reports exactly the slots that record labels: 304 of the 320 Change slots
(106 Edit, 76 Repair, 122 Consistency). Every overall column uses all 320, and
we report the two denominators separately rather than imputing a label. Per-capability and overall columns
therefore rest on different denominators, and we keep them distinct
throughout.
\end{minipage}\hfill
\begin{minipage}[t]{0.44\textwidth}
\vspace{0pt}
\small
\centering
\caption{Continuity facets.}
\label{tab:facet_inventory}
\setlength{\tabcolsep}{12pt}
\renewcommand{\arraystretch}{1.14}
\begin{tabular}{lr}
\toprule
\rowcolor{HeaderColor}
\textbf{Facet} & \textbf{Tasks} \\
\midrule
Design & 102 \\
Experience & 85 \\
Regression & 52 \\
Content/assets & 36 \\
State & 27 \\
Implementation & 15 \\
Unnormalized & 3 \\
\bottomrule
\end{tabular}
\end{minipage}
\end{table}

\section{Gate Families}
\label{app:gates}

The decomposition into completion and additional verification is computed from names
emitted by the frozen verifier, not from post hoc labeling.
Table~\ref{tab:gates} lists the families used in the reported analysis.
\texttt{visual\_regression} is held out of the conservative family throughout
because it compares against a reference render without controlling scroll
offset, so target-versus-prior provenance is not uniform across its failures.
One completion signal, \texttt{state\_url\_signals}, is configured by no task in
the locked suite and so passes on every cell; we keep it in the table because
the verifier emits it and Gate\% is divided by 19 throughout. Recomputing
Gate\% over the eighteen signals that can fail lowers every model by two to
three points and leaves the ordering in Table~\ref{tab:main_pwb_modb}
unchanged.

\begin{table}[h]
\centering
\small
\setlength{\tabcolsep}{8pt}
\renewcommand{\arraystretch}{1.25}
\caption{Deterministic gate families. The three partition all 19 gates.}
\label{tab:gates}
\begin{tabular}{>{\raggedright\arraybackslash}p{0.27\textwidth}
                >{\raggedright\arraybackslash}p{0.60\textwidth}}
\toprule
\rowcolor{HeaderColor}
\textbf{Family} & \textbf{Verifier signals} \\
\midrule
Completion (12) & completion text, visible text, DOM assertions, required
states, interaction assertions, state quality, action history, screenshots,
asset paths, exact asset paths, image-alt signals, state-URL signals \\
Conservative extra (6) & regression text, visual-anchor similarity, horizontal
overflow, console errors, forbidden text patterns, state artifacts \\
Broad sensitivity only (1) & visual regression \\
\bottomrule
\end{tabular}
\end{table}

\section{Presentation Interfaces and Evaluation Arms}
\label{app:interfaces}

Two interfaces read the same frozen package, and the paper uses both. They are
reported separately and are not comparable. Table~\ref{tab:arms} states where
they differ, so no result is read across the boundary.

\begin{table}[h]
\centering
\small
\setlength{\tabcolsep}{7pt}
\renewcommand{\arraystretch}{1.25}
\caption{The two evaluation arms, differing only in presentation and turn control.}
\label{tab:arms}
\begin{tabular}{>{\raggedright\arraybackslash}p{0.20\textwidth}
                >{\raggedright\arraybackslash}p{0.33\textwidth}
                >{\raggedright\arraybackslash}p{0.33\textwidth}}
\toprule
\rowcolor{HeaderColor}
& \textbf{Staged} & \textbf{Full-spec} \\
\midrule
Used in
& Tables~\ref{tab:completion_continuity}, \ref{tab:regimes},
  \ref{tab:main_pwb_modb} and Section~\ref{sec:results}
& Table~\ref{tab:cli_scaffold}, Figure~\ref{fig:appx_scaffold} \\
Solver input
& One work unit at a time, via the simulated user
& Complete task and repository context, once \\
Turn control
& Our driver: 12 turns per stage, 200 per run
& The agent's own loop, under a time and token ceiling \\
Feedback access
& Per-gate failure text each checkpoint, plus a screenshot on the VLM leg
& None; scored on what it submits \\
Protected-state replay
& Yes: cleared assertions re-verified at every later checkpoint
& No: one terminal verification \\
Metrics defined
& Task\%, Gate\%, Stage\%, WCS, breaks
& Task\%, Gate\%, component mean \\
\bottomrule
\end{tabular}
\end{table}

The simulated user is an evaluation interface, and deliberately so: one
capability-conditioned persona per task, revealing requirements authored in advance and
grounded feedback while acceptance stays fixed. That separation is what makes
runs comparable across thirteen models. Multi-role interaction is a natural
extension and belongs on the same frozen states and gates.

Scaffold and staged scores are therefore not comparable as
model-quality estimates. Gate\% in Table~\ref{tab:cli_scaffold} runs well above
Table~\ref{tab:main_pwb_modb} while strict Task\% runs below it, for reasons
that have nothing to do with agent quality: a full-spec agent is graded once,
against the final requirement set, without ever having to hold an earlier state
passing. The two \textbf{Task\%} columns are not the same quantity either: the
scaffold arm reports a final gate pass, the staged arm the stricter whole-task
notion of Section~\ref{sec:protocol}. That is precisely the
difficulty the staged interface adds, and why we read the scaffold study as an
internal comparison rather than a second leaderboard.

\section{Verifier Validity}
\label{app:verifier}

\textbf{Admission fixes the acceptance contract before any solver is run.} All
479 references pass every gate and all 479 authored bad solutions fail, and the
Build controls additionally pass the empty-baseline, self-certification,
distinct-artifact and stage-solvability checks. That is true by construction, because a task enters the pool only when its contract is
demonstrably satisfiable and its plausible failure mode is demonstrably caught,
and it is exactly the guarantee we need: before a solver runs, two construction
failures are already ruled out, a reference that cannot satisfy its own
contract, and a gate that waves the authored regression through.
Accuracy on unseen solver output is a separate question, and we
measure it directly on real sweep output below.

\textbf{Blind model audit on solver output.} We drew 720 stratified
(cell, gate) verdicts from the actual sweep and rejudged each one from evidence
alone. Rater A is the deterministic auto-rater; Rater B is ten blind shards of a
frontier model that see the captured DOM, screenshot, and console log for the
state in question, together with the assertion text, but never the gate
implementation, the reference patch, or the verifier's own verdict. On the
$n=359$ subset where both raters commit, they agree $94.2\%$ of the time at Cohen
$\kappa=0.820$. Rater A abstains by design on gates whose evidence lies outside
the blind bundle, so the committed subset is the informative one. Against the A/B consensus label ($n=338$) the frozen verifier
reaches $\kappa=0.766$, with a false-pass rate of $0.044$ [$0.025$, $0.075$] and
a false-fail rate of $0.190$. Stratified by capability, $\kappa$ is $0.862$ on
Edit ($n=114$), $0.791$ on Consistency ($n=117$), and $0.657$ on Repair
($n=107$); by family it is $0.772$ on completion ($n=240$) and $0.711$ on
continuity ($n=98$).

\textbf{Which side of the ledger each claim rests on.} The two error rates act
on our numbers differently, and the $19.0\%$ deserves care rather than
reassurance. Provenance answers whether a state was passing before, not whether
the later failure is correct. A false \emph{pass}
($4.4\%$) would put a run into the completing subset that does not belong there,
so it bears on the denominator of every post-completion figure, and that is
the tight side. A false \emph{fail} acts in opposite directions depending on
which gate produces it. On a completion gate it withholds completion from a run
that deserved it. That removes runs from the completing subset, and since the
removed runs need not regress at the rate the remaining ones do, it changes the
composition of the conditional sample in a direction we cannot sign: dropping
low-regression runs would raise the measured rate and dropping high-regression
runs would lower it. We therefore read every post-completion figure as
conditional on the verifier's completion verdict, not as a conservative bound.
The strict Task\% column of Table~\ref{tab:main_pwb_modb} ($1.25$ to $3.50\%$)
is a floor against false fails only; the $4.4\%$ false-pass rate acts on it in
the other direction. On a \emph{continuity} gate it would do real damage: a
state that still works, judged failed, is counted as a regression that did not
happen, inflating the $28.2\%$ gap and both provenance-exact channels.

\textbf{Where the false fails actually fall.} That distinction is measurable in
the audit sample. Of the 63 verifier-FAIL verdicts that reach blind consensus,
12 are false fails, the $19.0\%$ above, and \emph{all twelve sit on
completion-family gates}: \texttt{screenshot\_files} (5, a criterion
the judge read as vacuous because none was configured),
\texttt{asset\_path\_signals} (4), \texttt{dom\_assertions} (2) and
\texttt{visible\_text\_signals} (1). On the conservative continuity family that
carries the regression signal, 0 of 7 consensus FAIL verdicts are false
(Wilson-95 $[0\%, 35.4\%]$). Seven verdicts cannot bound that rate tightly, and
we do not claim they do. Gate family and temporal role are also different axes.
A \texttt{dom\_assertions} check belongs to the completion family while its
stage is being cleared and then joins the protected set, so a false fail on it
at a later checkpoint would be recorded as a protected-state regression even
though this audit files it under a completion-family gate. The unit audited here
is the family of a verdict, not the regression event, and it therefore cannot on
its own bound the error in the $10.8\%$. Sampling recorded regression events and
checking the before-and-after evidence for the same assertion is the audit this
needs, and the first extension we would run.

\textbf{What carries the class-(a) regressions.} The same holds for the
events themselves. Every one of the 66 class-(a) regressions on the
covered completing subset rests on
\texttt{no\_horizontal\_overflow} (37 events) or \texttt{console\_errors} (32),
and none rests on a gate of the capture-shaped kind
(\texttt{required\_states}, \texttt{state\_artifacts},
\texttt{screenshot\_files}). Those two gates are narrow measurements: whether the document scrolls sideways
at a fixed viewport, and whether the browser console throws. The same gate
implementations produce both verdicts, but the initial baselines are captured in
our analysis environment and the final workspaces come from the evaluation
cluster, so browser build, dependency versions and capture timing are matched by
construction only up to that boundary. Appendix~\ref{app:reproducibility}
reports what we can say about the two environments agreeing, on a sample small
enough that we read the number as supported rather than established. The capture-shaped failure categories visible in
Figure~\ref{fig:pristine_regression} belong to the $43.9\%$ view over all runs,
not to the completing subset the $18.3\%$ is computed on. In panel (b) the two
largest entries mean the final workspace could no longer be loaded or captured
at the required state, which is the most severe outcome rather than a
measurement artifact, and \texttt{visual\_regression} is excluded throughout for
the same reason it leaves the conservative family; including it would move the
pooled rate by under half a point in either direction.

\begin{figure}[t]
\centering
\includegraphics[width=\linewidth]{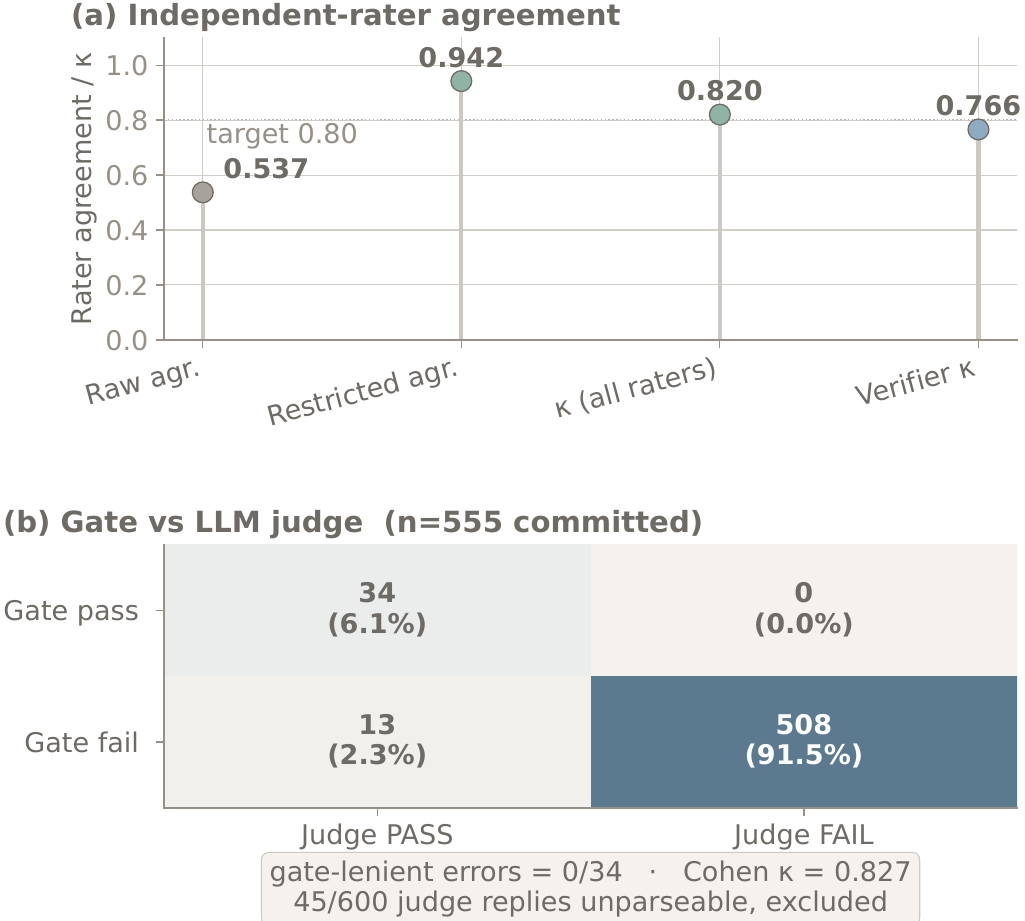}
\caption{Verifier calibration. Every one of the 479 references passes all gates
and every one of the 479 authored bad solutions fails. (a) Agreement and Cohen
$\kappa$ against ten blind raters. (b) Agreement with a strong LLM judge over
600 verdicts.}
\label{fig:appx_verifier}
\end{figure}

\textbf{Independent-judge cross-check.} As a second, cheaper check we sent
600 cells stratified across lh-score buckets and 14 models to a strong LLM judge
with the same evidence bundle and no access to gate internals. Forty-five replies
(7.5\%) could not be parsed into a verdict; they are excluded rather than folded
into the denominator, leaving $n=555$ committed verdicts. On those, agreement is
$97.7\%$ with Cohen $\kappa=0.827$. The error is one-sided in the reassuring
direction: the judge never overturned a gate PASS in any of the 34 cases, and all 13
disagreements are cells the gate failed and the judge would have passed
(Figure~\ref{fig:appx_verifier}b). Because the parse failures are not verdicts,
they cannot be scored either way, so we report both the $555$ denominator and
the 45 exclusions.

\begin{figure}[t]
\centering
\includegraphics[width=\linewidth]{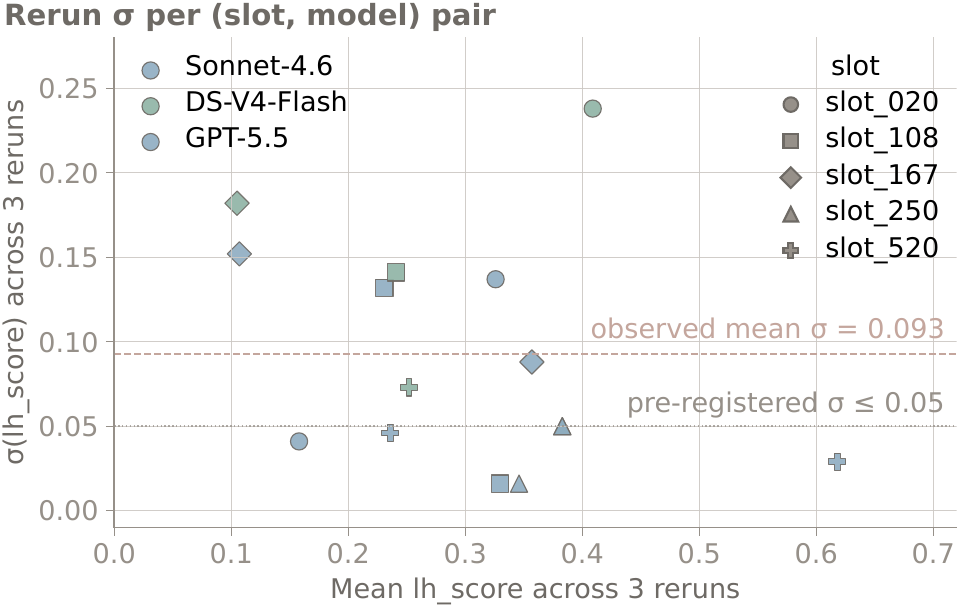}
\caption{Rerun stability at temperature zero: fifteen pairs of one slot and one model replayed
three times, plotting the standard deviation of lh\_score against its mean. The
mean standard deviation of 0.093 is the noise floor on every per-cell graded
score, and the binary verdict is unchanged in all fifteen pairs. Replays
re-execute the whole solver loop, so this is end-to-end run variability.}
\label{fig:appx_determinism}
\end{figure}

\textbf{A measured noise floor.} Fifteen pairs of one slot and one model replayed three times
at temperature zero give a mean lh\_score standard deviation of 0.093, above our
pre-registered target of 0.05, while the binary verdict is stable in all fifteen
pairs. We therefore attach this floor to every per-cell graded number.
Aggregate contrasts are judged by their own paired confidence intervals, not
against this per-cell spread, which is a different quantity on a different
metric. Because
a replay runs the entire solver loop again, the observed $\sigma$ in
Figure~\ref{fig:appx_determinism} measures end-to-end run variability, and it
does not isolate the verifier. The gate predicates are deterministic functions
of the captured evidence, but the capture itself drives a browser with
asynchronous loading, fonts and a dev server, so a deterministic rule does not
by itself make the measurement stable. Replaying the verifier alone against
fixed initial, passing and failing artifacts, and reporting the gate-flip rate,
is a separate experiment we have not run; it is the most direct remaining check
on spurious pass-to-fail transitions, and we do not claim its result in this
version of the paper.

\section{State Provenance and the Regression Estimate}
\label{app:regression}

\textbf{Three classes of state.} A Change run touches states of three kinds:
(a) states already working in the initial repository and never named by the task;
(b) states the run itself established by clearing an earlier stage; and (c) the
target state of the current stage. Only (c) is covered by the completion family.
The paper's two continuity numbers cover (a) and (b) very differently, and
conflating them is the main interpretive hazard in the result.

\textbf{What 28.2\% is.} The continuity checks -- the conservative
non-completion family (console errors,
forbidden text, horizontal overflow, regression text, state artifacts,
visual-anchor similarity) fires on the final workspace. A cell in that family may
be breaking a class-(a) state, a class-(b) state, or simply failing a global
hygiene check that was never passing in the first place; the emitted artifacts do
not record which. The $190/674=28.2\%$ figure is therefore a
\emph{post-completion verification gap}, the rate at which finishing the
request fails to imply passing everything else we check, and is reported as
such everywhere in the paper. Only 35 of those 190 also record a
protected-state regression, so the gap is largely not the replay channel
restated. Its Wilson-95 interval is [24.9\%, 31.7\%] and its
task-clustered bootstrap interval [19.2\%, 38.3\%]; per-model point estimates
run from 12.1\% to 38.2\%, and no model is free of the gap. The broad definition adds
\texttt{visual\_regression}, which is excluded from the conservative family for
exactly this reason: it compares against a reference render without recording
whether the differing region belongs to the target or to the incumbent page.

\textbf{What 10.8\% is.} Class (b) is mechanically identifiable. When stage $k$
clears, the harness appends its acceptance assertions to a protected set, and
every subsequent checkpoint in that run re-verifies the whole protected set
before judging the new stage. Because the whole sequence is authored and frozen
before any solver runs, a later stage that legitimately replaces earlier wording,
ordering or state is authored so that the superseded assertion is not in the
protected set: admission requires the reference implementation to pass every
protected assertion at every later checkpoint, so a plan in which a later
requirement contradicts an earlier one fails admission and never enters the
suite. Protected assertions therefore stay in force for the rest of the run by
construction, and the check on that condition is the reference replay rather
than a promise about the authoring. A protected assertion that passed at stage $k$ and
fails at stage $k'>k$ is a pass-to-fail transition on a state the run established
itself, recorded at the moment it happens; no post hoc labelling is involved. The
solver is told about it explicitly and given a chance to repair it, which is why
the event is worth counting on its own.

\begin{table}[h]
\centering
\small
\setlength{\tabcolsep}{8pt}
\caption{Protected-state regressions and the post-completion gap only partly
coincide. Rows: whether the run recorded at least one protected-state
pass-to-fail transition. Columns: whether the final workspace fails a
conservative non-completion check. All 674 completing Change cells.}
\label{tab:regression_crosstab}
\begin{tabular}{lrrr}
\toprule
\rowcolor{HeaderColor}
& \textbf{Conservative fail} & \textbf{Conservative clean} & \textbf{Total} \\
\midrule
Protected-state regression & 35 & 38 & \textbf{73} \\
No regression recorded     & 155 & 446 & \textbf{601} \\
\midrule
\textbf{Total} & \textbf{190} & \textbf{484} & \textbf{674} \\
\bottomrule
\end{tabular}
\end{table}

Table~\ref{tab:regression_crosstab} makes the relationship concrete. The two
measures agree on 35 cells. The 155 cells that fail conservatively without a
recorded regression are the ones protected-state replay alone cannot place: some
are genuine class-(a) regressions the harness never had a chance to observe
passing, others are hygiene failures that were always failing. The 38 cells in
the other off-diagonal regressed a protected state mid-run and repaired it before
the final verdict, which the final-state gap cannot see at all. Neither number
subsumes the other, and reporting only one of them would misstate the result.

\textbf{Class (a), measured.} The initial repository is capturable, so class
(a) does not stay hypothetical. Running the frozen suite on the repository as
handed covers 234 slots. Of the completing cells that fail a
conservative non-completion check without a recorded protected-state
regression and a capturable baseline (126 of the 155), 50 fail a gate the
untouched repository passed, class-(a) regressions the harness could
not previously certify, and 76 fail a gate that was already failing before the
run, i.e. a hygiene property the task never fixed. The other 29 sit on slots
whose repository does not render as handed, so they have no baseline. Provenance
is therefore resolved for all three classes: across the 586 completing cells
with an untouched-repository baseline, the two channels identify 107
provenance-exact regressions, 18.3\% rather than the 9.7\% the protected set
alone reaches; over all
completing runs the protected-state rate is unchanged at $73/674=10.8\%$.

\textbf{Coverage of the class-(a) baseline.} A slot contributes a baseline only
if the repository as handed renders offline under the frozen toolchain: 234 of
the 320 Change slots qualify, and 128 of those are served by the project's own
dev server rather than as a built static tree. The 86 excluded slots skew towards
stacks with a build step (58\% versus 41\% among included slots), though
no-build static sites are the second-largest excluded group, so the exclusion is
not specific to bundlers. Two checks bound the effect. The protected-state rate
on the covered completing subset is $9.7\%$ against $10.8\%$ on all 674
completing cells, so no shift of that size is detectable between them;
and \texttt{visual\_regression} is excluded from the class-(a) definition for the
same reason it is excluded from the conservative family, which moves the pooled
rate by $0.2$ points.

\textbf{Robustness of the estimate.} The rate is $73/674=10.8\%$ [$8.7\%$,
$13.4\%$] among completing runs and $823/6{,}305=13.1\%$ [$12.2\%$, $13.9\%$]
among non-completing runs; the intervals overlap, so no difference of
that size is detectable between them, which is a bound on the association,
not evidence that none exists. Per-model point estimates span
$3.6\%$ (MiniMax-M3) to $20.7\%$ (GPT-5.5) over 32 to 80 completing cells each, and
the ordering does not track overall WCS: the two strongest models by WCS sit at
$6.2\%$ and $20.7\%$ respectively. Regression propensity is thus a partly
independent axis of model behaviour, which is the case for measuring it. Two
caveats hold for every cross-model comparison of this rate. Each model is
conditioned on the tasks it completed, and those sets differ; and a model that
clears more stages is verified at more checkpoints, so it has more opportunity
for a protected assertion to be observed flipping. We therefore read the
per-model rates as descriptive, and compare models only like for like on their
commonly completed tasks.

\begin{table}[h]
\centering
\small
\caption{Provenance-exact regression by observation point, in full; the body
carries the headline rows as Table~\ref{tab:regression_endpoints}. The endpoint
rows are run-level: they ask whether \emph{any} conservative gate fails at the
end, not whether the specific assertion that flipped was repaired. Brackets:
Wilson-95; clusters are tasks.}
\label{tab:regression_endpoints_full}
\setlength{\tabcolsep}{8pt}
\renewcommand{\arraystretch}{1.14}
\begin{tabular}{>{\raggedright\arraybackslash}p{0.46\linewidth}rll}
\toprule
\rowcolor{HeaderColor}
\textbf{Event} & \textbf{Count} & \textbf{Rate [Wilson-95]} & \textbf{Clustered-95} \\
\midrule
\multicolumn{4}{@{}l}{\textit{Protected-state replay (all completing Change runs)}} \\
Regression at any checkpoint & 73/674 & \textbf{10.8\%} [8.7, 13.4] & [6.8, 15.6] \\
\quad final conservative checks clean & 38/674 & 5.6\% [4.1, 7.6] & [3.0, 8.8] \\
\quad a conservative check still fails & 35/674 & 5.2\% [3.8, 7.1] & [2.3, 8.9] \\
\midrule
\multicolumn{4}{@{}l}{\textit{Both channels (covered completing subset)}} \\
Class (a): initial state, final failure & 66/586 & 11.3\% [9.0, 14.1] & [5.8, 18.2] \\
Class (b): regression at any checkpoint & 57/586 & 9.7\% [7.6, 12.4] & [5.8, 14.6] \\
\quad counted by both channels & 16/586 & 2.7\% [1.7, 4.4] & [0.5, 6.5] \\
Union of the two channels & 107/586 & \textbf{18.3\%} [15.3, 21.6] & [12.3, 25.4] \\
\quad a conservative check still fails & 78/586 & \textbf{13.3\%} [10.8, 16.3] & [7.6, 20.4] \\
\bottomrule
\end{tabular}
\end{table}

\textbf{Trajectory against endpoint.} Replay fires whenever a cleared assertion
flips, and 38 of the 73 runs that record one finish with every conservative
check clean, so the trajectory rate is roughly twice the end-state rate. The
untouched-repository channel is a final-state measure by construction, which is
why the two are reported at both observation points rather than merged.

\textbf{The protected-state endpoint is not yet resolved by these arms.} Applying
the same protected-state test to every single-variable arm we ran gives no
significant movement at the available sample sizes. Table~\ref{tab:arm_endpoint}
lists the counts: no arm shifts more than a handful of cells, and no paired
statistic clears its own standard error. The gate-level effects reported in
Section~\ref{sec:results} are real; the claim that they operate
\emph{through} the regression channel is not supported by these data, and we do
not make it. Arms powered for this endpoint are the natural next experiment.

\begin{table}[h]
\centering
\small
\caption{Cells recording a protected-state regression, before and after each
single-variable arm.}
\label{tab:arm_endpoint}
\setlength{\tabcolsep}{9pt}
\renewcommand{\arraystretch}{1.12}
\begin{tabular}{lccr}
\toprule
\rowcolor{HeaderColor}
\textbf{Arm} & \textbf{Baseline} & \textbf{After} & \textbf{Paired $t$} \\
\midrule
History truncated to five turns & 0/14 & 2/14 & $+1.07$ \\
Preservation clause added & 12/99 & 7/99 & $-0.57$ \\
Edit emission, u-diff & 5/20 & 6/20 & $-0.42$ \\
Edit emission, full-file & 5/20 & 2/20 & $-0.98$ \\
Simulated user always sees a screenshot & 0/13 & 1/13 & --- \\
Simulated user sends error text only & 0/14 & 3/14 & --- \\
\bottomrule
\end{tabular}
\end{table}

\section{Additional Result Detail}

\label{app:results}

\textbf{A worked WCS, and what the suite counts are.} One cell makes
Equation~\ref{eq:wcs} concrete. Slot 1 under Kimi-K3, text-only leg, clears 3 of
its 5 stages ($s=0.600$), passes 10 of 19 gates ($g=0.526$), so does not pass
every gate on its final workspace ($v=0$), and records no regression break
($b=0$), giving
$0.5(0.600)+0.3(0.526)=0.458$, the value in the released manifest. Note that
$s$ counts stages only: this run also has 4 of 7 \emph{units} cleared once
reworks are included, which is the separate completion rate we report, not the
$s$ of the score. Because $b_i$ accumulates once per (checkpoint, gate),
the penalty is sensitive to how often a run is verified: $0.05b_i$ exceeds
$0.2$ on $576$ of the $8{,}737$ cells, where the clamp then floors the score at
zero.

The task counts in the paper refer to different sets and are not
interchangeable. The validated authoring pool holds 479 tasks, each with a
passing reference and a failing authored control. The locked evaluation suite is
the 400 of those used for every number here, 320 Change and 80 Build. Of the 320
Change tasks, all 320 rebuild from the frozen package in the analysis
environment and 234 of those also serve a capturable initial baseline, which is
the coverage of the class-(a) channel; 128 of the served ones run the project's
own dev server rather than a built static tree. A
further 16 Change slots carry no capability label in the frozen manifest and
enter overall columns only. Task-clustered bootstraps resample tasks with all of
their cells, so repeated runs of one task are not treated as independent. Tasks
derived from the same upstream repository could still correlate beyond that, and
we do not currently report a repository-level clustering; the released manifest
carries the source path for every task, so that grouping can be recomputed from
the release without rerunning anything.

\textbf{Run configuration.} Solver and simulated user are separate models. The
simulated user is held fixed across the sweep so that persona behaviour is not a
free variable between models, and the persona names in
Appendix~\ref{app:interfaces} label prompt variants rather than model
configurations. Every cell records, in its released row, the solver
model identifier, the leg, the number of turns and tokens
actually used, the edits applied and rejected, the per-stage trace, and which
gates failed; the route, viewport and action prefix each gate was evaluated at
is fixed per state by the released \texttt{state\_plan.json} of that task. Infrastructure
failures (install failure, timeout, provider error) are written as
\texttt{infra\_failed} and the cell is re-run rather than scored, so they do not
enter any denominator; a gate that a task does not configure is not counted in
that task's gate total, which is why gate totals vary by task and why Gate\% is
a rate rather than a count. The five \texttt{screenshot\_files} false fails in
the audit sample are exactly the case where a judge read an unconfigured
criterion as vacuously failing, and they are counted against the verifier in
Appendix~\ref{app:verifier} rather than excused.

\textbf{Models, denominators and statistical conventions.} The 13 models are
Claude Opus 4.8 and Sonnet 4.6, GPT-5.5, Gemini 3.5 Flash, DeepSeek V4 Pro and
Flash, GLM-5.2-FP8, Hunyuan 3 Preview, Kimi K2.7 Code and K3, MiniMax M3, Qwen
3.6 and Qwen 3.7 Max. The four without a screenshot leg send no image on any
screenshot cell, so they are evaluated text-only rather than simulated with
empty image turns, and a cell scored against a superseded specification is
rerun rather than carried forward. Denominators differ by question: the
post-completion gap uses all 6,979 Change cells, the long-horizon analysis the
1,758 Build cells stratified by stage count, the Build against Change comparison 7,142
cells from the nine dual-modal systems, and the modality experiment 3,558
within-model, within-task pairs from those same nine. The post-completion gap
and both regression channels resample tasks with all of their cells, because
several runs of the same task differ only in model or modality. Paired modality differences use
Wilcoxon signed-rank tests; paired arms are judged by their own confidence
intervals. Long-horizon results are bucketed by stage count, Kruskal--Wallis
diagnostics are exploratory, and Benjamini--Hochberg correction is applied
throughout to the framework and the facet families.

\begin{table}[h]
\centering
\begin{minipage}[t]{0.52\textwidth}
\vspace{0pt}
\small
\centering
\caption{Headline diagnostics with their eligible sets. Each row uses its own
denominator, stated in the $N$ column.}
\label{tab:diagnostics}
\setlength{\tabcolsep}{4pt}
\renewcommand{\arraystretch}{1.12}
\begin{tabular}{>{\raggedright\arraybackslash}p{0.54\linewidth}rr}
\toprule
\rowcolor{HeaderColor}
\textbf{Diagnostic} & $N$ & \textbf{Result} \\
\midrule
Reference passes & 479 & 100.0\% \\
Smart-wrong fails & 479 & 100.0\% \\
Conservative gap & 674 & \textbf{28.2\%} \\
Broad sensitivity & 674 & 64.1\% \\
Applied $>$ rejected & 8{,}539 & \textbf{82.0\%} \\
Interaction fails, DOM ok & 1{,}322 & \textbf{29.5\%} \\
Radius $\geq2$ & 432 & 31.9\% \\
\bottomrule
\end{tabular}
\end{minipage}\hfill
\begin{minipage}[t]{0.44\textwidth}
\vspace{0pt}
\small
\centering
\caption{Regime comparison over 7,142 dual-modal cells (1,439 Build, 5,703
Change). The WCS gap tracks stage completion, not gate quality.}
\label{tab:regimes}
\setlength{\tabcolsep}{9pt}
\renewcommand{\arraystretch}{1.12}
\begin{tabular}{lrr}
\toprule
\rowcolor{HeaderColor}
\textbf{Metric} & \textbf{Build} & \textbf{Change} \\
\midrule
$N$ & 1{,}439 & 5{,}703 \\
WCS & 0.240 & 0.348 \\
Stage pass & 0.210 & 0.411 \\
Gate pass & 0.525 & 0.543 \\
Stages & 8.7 & 6.3 \\
\bottomrule
\end{tabular}
\end{minipage}
\end{table}

\textbf{The failures are not patch machinery.} The 8,737 live cells divide
exactly into the 198 whole-task passes of Table~\ref{tab:main_pwb_modb} and 8,539
failures. Among the failures, 8,396 (98.3\%) land at least one edit; in 7,002
(82.0\%) the cumulative count of applied edits exceeds rejected edits and the
run still does not pass; 1,394 (16.3\%) edit the workspace but have rejected
edits at least as numerous as applied ones; and only 143 (1.7\%) land no edit
at all, so a no-edit explanation covers under two per cent of them. Landing one
edit does not exclude a later patch failing on format or scope, so these counts
bound the tooling account rather than eliminating it.

\begin{figure}[h]
\centering
\includegraphics[width=0.98\linewidth]{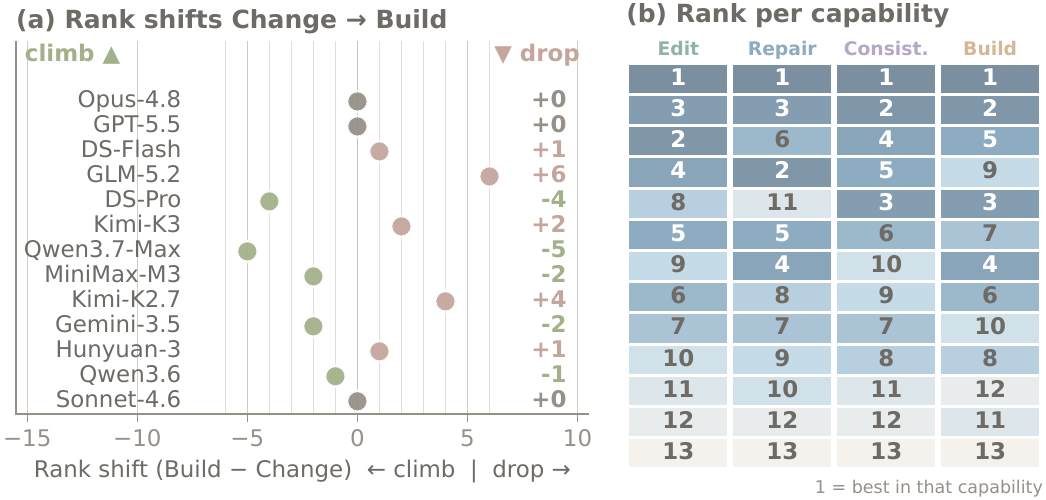}
\vspace{-6pt}
\caption{(a)~Rank shift from Change to Build per model; negative is a climb.
(b)~Each model's rank within each of the four capabilities, with 1 the best
rank in that column.}
\label{fig:build_reshuffles}
\end{figure}

\textbf{Ranking coherence across capability slices.} The mean Spearman
correlation between a per-capability WCS ranking and the overall WCS ranking is
0.885:
Edit $\rho=0.934$, Consistency $\rho=0.934$, Build $\rho=0.852$, Repair
$\rho=0.819$. Agreement on the overall order is thus strong on every slice, and
Build is not the weakest of them; what distinguishes Build is that it permutes
the top of the ranking rather than the whole of it
(Figure~\ref{fig:build_reshuffles}).

\begin{figure}[h]
\centering
\includegraphics[width=0.98\linewidth]{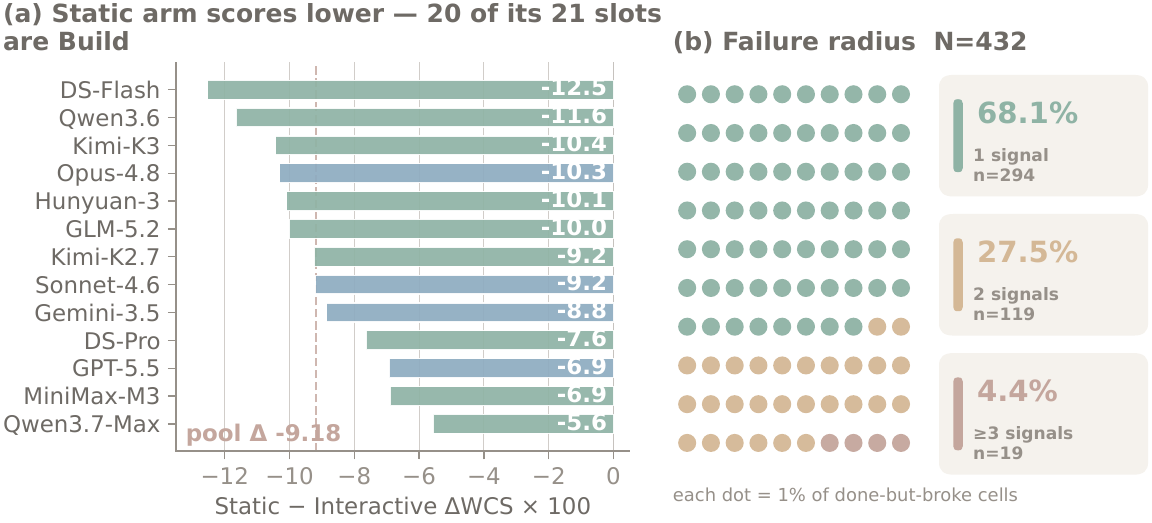}
\vspace{-7pt}
\caption{Interactive-vs-static $\Delta$WCS per model and cumulative failure breadth.}
\vspace{-4pt}
\label{fig:interaction_and_radius}
\end{figure}

\textbf{The static arm is not a comparable half.} Classifying the suite by
whether a task's frozen specification declares interaction assertions splits it
into 379 interactive slots and 21 static ones. The pooled within-model gap is
$-9.18$ points of $100\times$WCS, static minus interactive
(Figure~\ref{fig:interaction_and_radius}a), but the split is confounded with
regime: 20 of the 21 static slots are Build slots, so the comparison is mostly
Build against Change. Within regime the gap collapses to $-0.90$ points on
Build and $-0.91$ on Change, and it is no longer negative for every model.
The slot-level split therefore does not carry an interaction result on its own.
The confound-free view is the paired one of Section~\ref{sec:results}: on
interactive cells whose static DOM requirement already passes, the interaction
assertions still fail 29.5\% of the time. What a static DOM assertion accepts
and what an action sequence accepts are not the same bar.

\textbf{Failure radius.} Within the 432 complete-but-not-cleared cells of the
broad visual-inclusive definition, a cell fails 1.37 distinct non-completion
signals on average: 31.9\% fail at least two and 4.4\% at least three. The count
is over individual signals, not over the three gate families of
Appendix~\ref{app:gates}. These are broad sensitivity diagnostics rather than
confirmed regression radii, since the broad family does not separate target
from prior state (Figure~\ref{fig:interaction_and_radius}b).

\begin{figure}[h]
\centering
\includegraphics[width=0.98\linewidth]{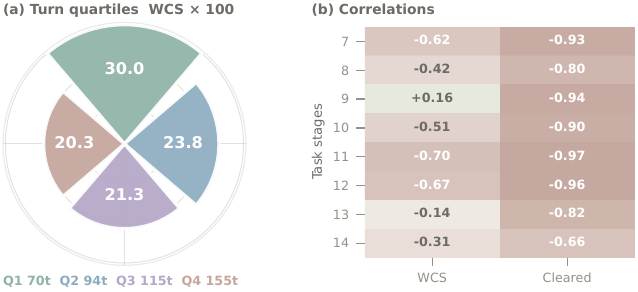}
\vspace{-8pt}
\caption{Turn-thrash on the long-horizon regime. (a) WCS by turn quartile.
(b) Correlation of turn count with WCS and with stages cleared, per
stage-count bucket.}
\label{fig:turn_thrash}
\end{figure}

\textbf{Model separation and the twin set.} The graded WCS has model-level
dynamic range on Build: an exploratory Kruskal--Wallis diagnostic gives
$H=118.6$ and $p=1.2\times10^{-19}$; because task cells repeat across models and
modalities we read it as a separation check rather than population-level
inference. On the greenfield twins of Section~\ref{sec:results}, stage counts
match by construction (twin mean 5.67 against source mean 5.81,
$\Delta=-0.14$), and eighteen of the twenty-one sit on no-build static or
vanilla-node sites. Restricting to the 70 pairs whose stage counts are
\emph{identical} leaves $\Delta\text{Gate\%}$ at $-7.67$, so the regime gap is
not an artifact of plan length. The paired comparison is therefore read within that regime,
where it holds the end state, the acceptance gates and the stage count fixed.

\textbf{The component mean, and what the grid does not control.} The scaffold
arm verifies once at the end, so WCS is undefined for it and the contrasts use
the \textbf{component mean}: the unweighted mean of the eight per-cell component
scores the verifier emits (change completion, the six continuity facets, and
browser execution), which needs no intermediate checkpoint and is defined in
both arms. The comparison is limited to two backbones and four shared
scaffolds, and we did not equalise time or token budgets across scaffolds, so
any budget difference sits inside the measured scaffold effect rather than being
controlled out of it.

\textbf{Scaffold versus backbone, in full.} On the matched $4\times2$ grid the
scaffold spread under a fixed backbone is $8.06$~pt on Sonnet-4.6 and
$10.52$~pt on GPT-5.5, against $5.08$~pt for the backbone swap under a fixed
scaffold; the unit is the per-cell component mean and the swap is paired by
task. The paired backbone swap favours GPT-5.5 for every scaffold, from $+0.60$
to $+8.27$~pt.
Scaffold rank order is largely preserved across backbones (Spearman
$\rho=0.80$), so the two factors are close to additive. Widening to all seven
Sonnet-4.6 scaffolds gives a mean pairwise component-mean gap of $5.68$~pt,
matching the backbone effect across a full model change.

\textbf{Change capability scores.} Across all 13 models, pooled Change WCS is
0.342 for Edit (2,310 cells), 0.450 for Repair (1,653), and 0.305 for Consistency
(2,665); a further 351 cells sit on 16 slots with no capability label in the
frozen manifest, excluded from this breakdown. Consistency remains 0.035
WCS below Edit ($\mathrm{se}=0.007$, $t=-5.07$, $p=4.0\times10^{-7}$) after
controlling for five task-size proxies and model fixed effects in a single linear
model on the 6,628 labelled cells; the category block itself is significant
(partial $F=182.48$ on $(2, 6608)$ degrees of freedom,
$p=7.3\times10^{-78}$). Every size-proxy Spearman correlation with WCS has
magnitude below 0.055, so category differences are not explained by specification
size alone. The size proxies are counts of required states, hidden states,
required content items, suggested files, and stages; no literal diff-line count
exists in the frozen artifacts, which is why proxies are used.

\begin{figure}[t]
\centering
\includegraphics[width=\linewidth]{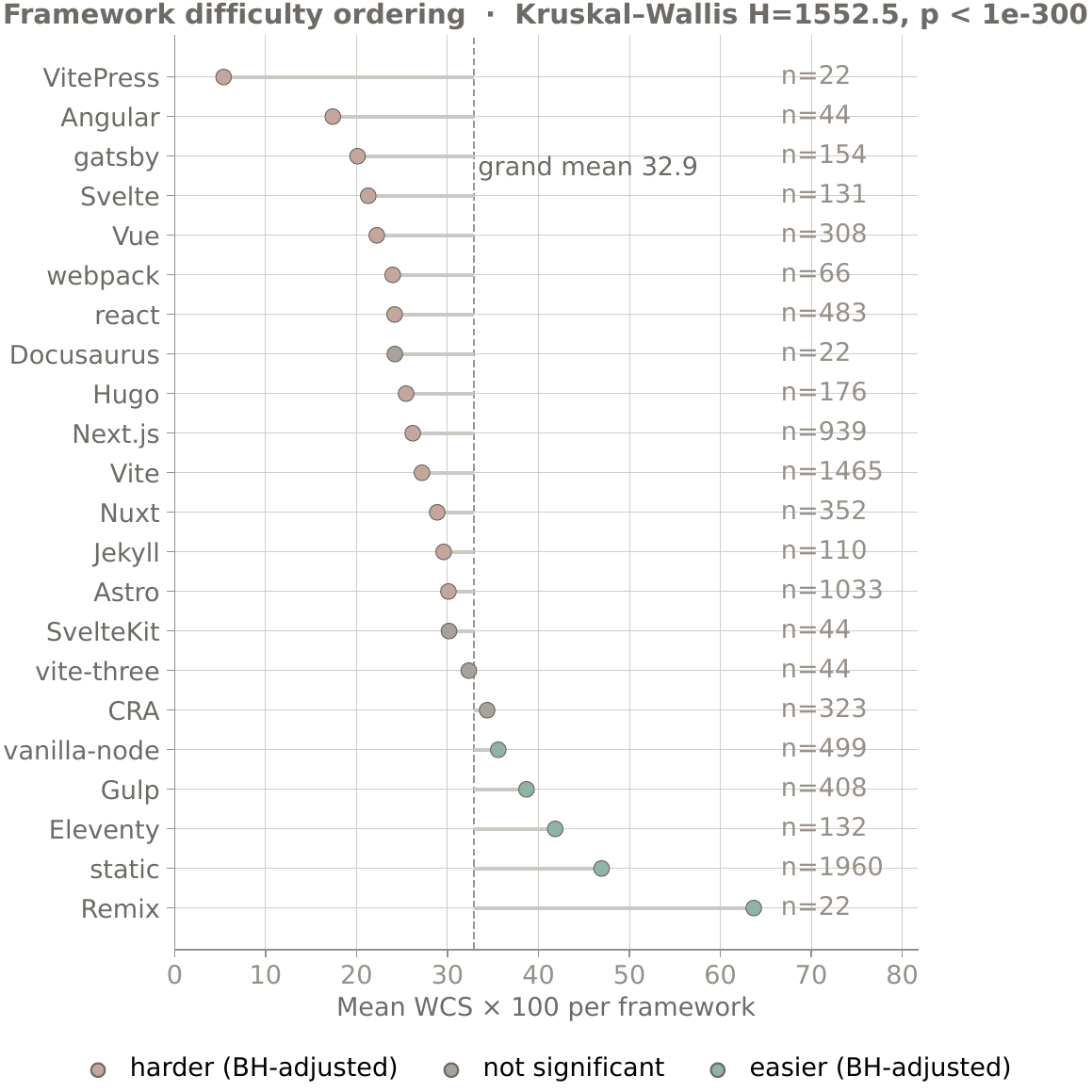}
\caption{Per-framework WCS ordering. Bars are Benjamini--Hochberg-adjusted harder (red) or easier
(green) than the grand mean; grey is not significant.}
\label{fig:appx_framework}
\end{figure}

\begin{figure}[t]
\centering
\includegraphics[width=\linewidth]{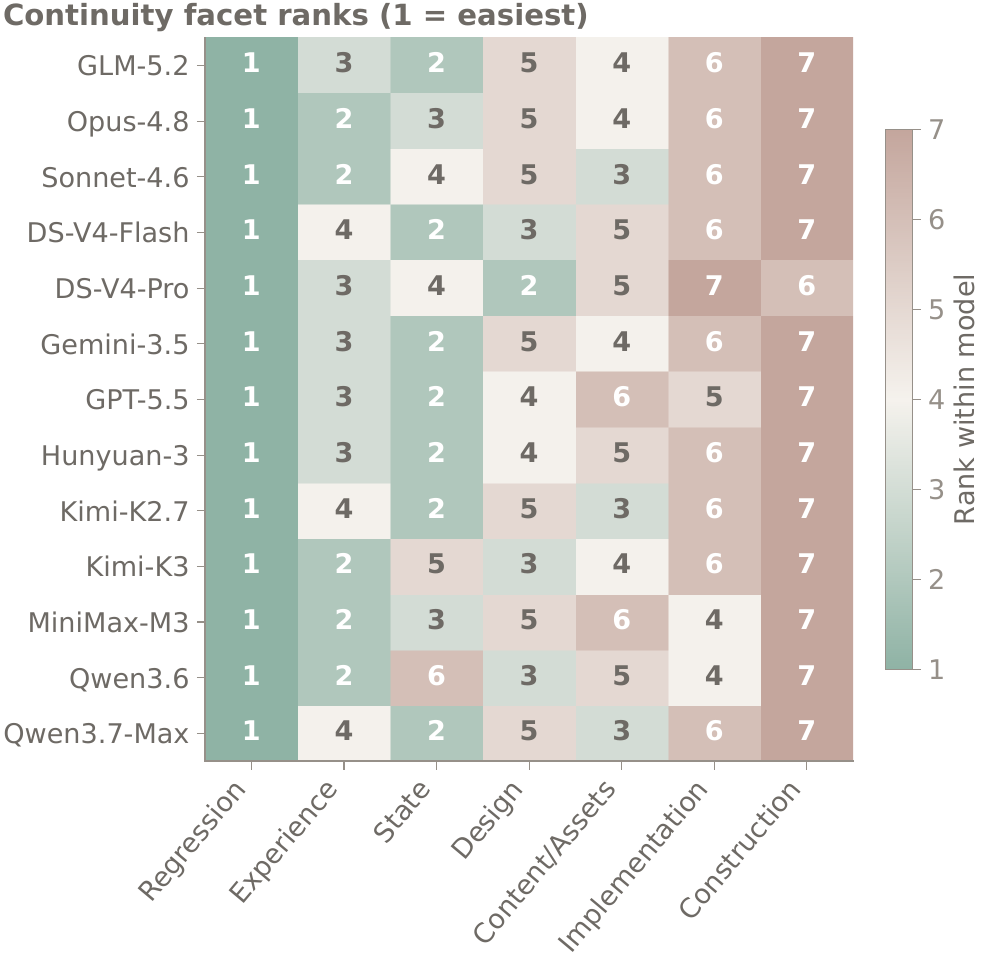}
\caption{Per-model facet ranking (1 = easiest for that model). All 13 models are
plotted and every facet is populated for every model. Regression-continuity is
the easiest facet for all 13; construction is the hardest for 12 of them.}
\label{fig:appx_facets}
\end{figure}

\begin{figure}[t]
\centering
\includegraphics[width=\linewidth]{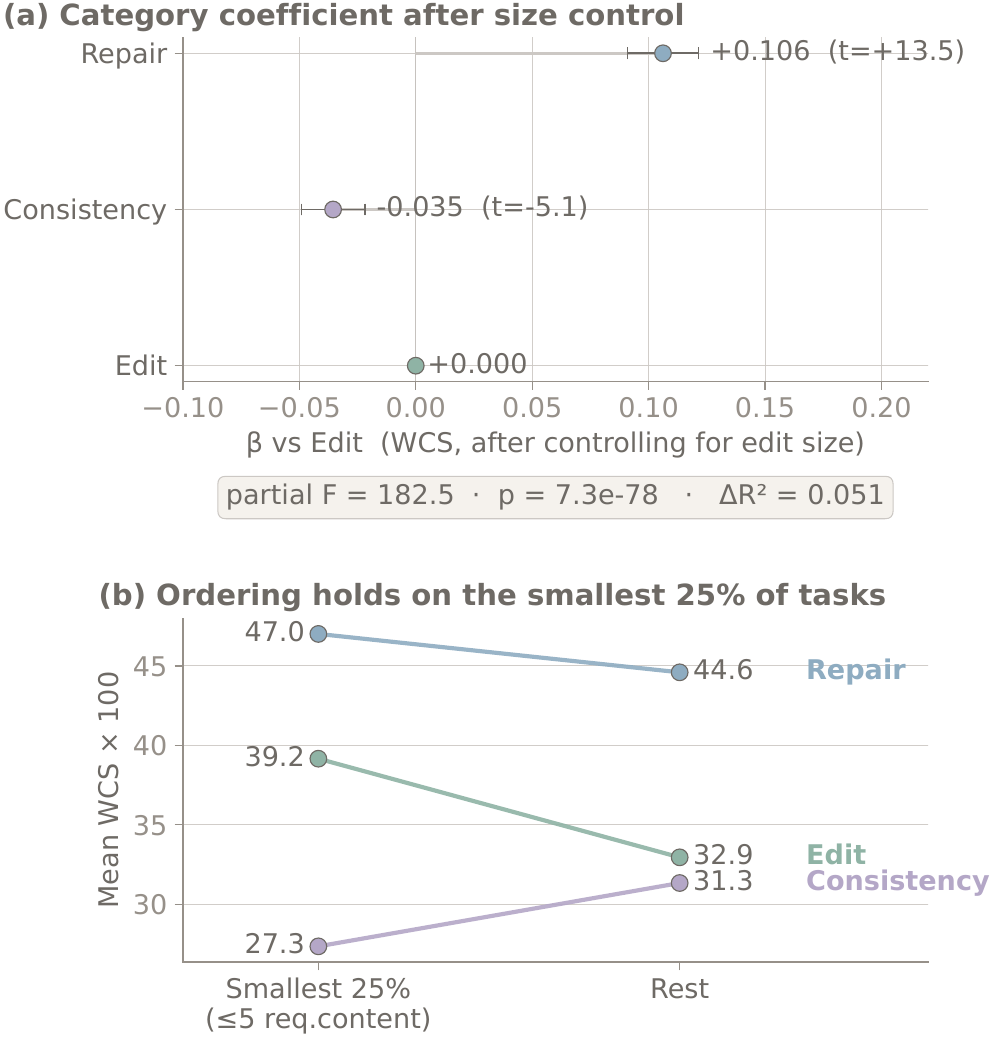}
\caption{Category effect survives edit-size control. (a)~Category coefficient
against the Edit baseline, controlling for task size and model. (b)~Mean WCS by
category on the smallest quarter of tasks, where the ordering is unchanged.}
\label{fig:appx_size}
\end{figure}

\textbf{Framework, facet, and size controls.} Three further breakdowns support
the category result in the main text. Per-framework WCS
(Figure~\ref{fig:appx_framework}) separates harder and easier stacks after
Benjamini--Hochberg adjustment; at the exploratory cell level the spread across
framework labels is large ($H=1552.5$ on 21 degrees of freedom, $p<10^{-300}$), with SPA
and bundler-heavy stacks at the hard end and static or lighter systems at the
easy end, though cells are not independent within a task so we read this as a
description of the observed panel rather than population inference. Per-model facet ranking
(Figure~\ref{fig:appx_facets}) shows regression-continuity is the easiest facet
for all 13 models and construction the hardest for 12 of them, so the facet
ordering is a property of the tasks rather than of any one model. The size control (Figure~\ref{fig:appx_size}) repeats the category
comparison on the smallest quartile of tasks and recovers the same ordering.

\textbf{Long-horizon boundary.} The negative relation between turns and progress appears only on long tasks.
Within five- and six-stage buckets, the correlation between turns and stages
cleared is positive (0.32 and 0.24 respectively). It becomes negative for every
bucket from 7 through 14 stages: $-0.93$, $-0.80$, $-0.94$, $-0.90$, $-0.97$,
$-0.96$, $-0.82$, and $-0.66$. This is why the main text reports a
within-length median rather than the pooled correlation. Failed units can consume
the whole budget, so the buckets diagnose cap persistence rather than the
causal effect of adding turns.

\textbf{Feedback-modality validity.} We call a model dual-modal only when more
than half of its screenshot-feedback cells actually send at least one image, which
guards against the failure mode of a text-only endpoint silently accepting and
discarding the image turn. Nine of the thirteen models clear this bar, with
image-carrying fractions from 0.807 to 0.912: Opus-4.8, Sonnet-4.6, Gemini-3.5,
GPT-5.5, Hunyuan-3, Kimi-K2.7, Kimi-K3, MiniMax-M3, and Qwen3.6. Four are excluded
because their screenshot leg sends images on \emph{zero} cells (GLM-5.2-FP8,
DeepSeek-V4-Flash, DeepSeek-V4-Pro, and Qwen3.7-Max), and are reported as not
applicable rather than simulated with empty image turns. The paired estimate
therefore rests on nine models and 3,558 within-model, within-task pairs; the
count exceeds the 400-task suite because each task contributes one pair per
qualifying model. The contrast is between assignment to the screenshot policy
and assignment to the text-only policy: image-carrying fractions run from 0.807
to 0.912 even among qualifying models, so a pair differs in the policy applied
rather than in whether the solver certainly saw an image on every turn.
Restricting to pairs that did carry an image would condition on run progress and
is not an unbiased comparison. On those pairs the mean change in WCS with
screenshot feedback is $-0.0007$, bootstrap interval [$-0.0064$, $0.0050$],
Wilcoxon $p=0.952$. The
completion proxy moves $+0.0037$ ($p=0.334$) and the broad visual-inclusive
continuity proxy $-0.0124$ ([$-0.0216$, $-0.0031$], $p=0.009$). Screenshots
help one facet, experience, by $+0.0144$ over $n=728$ pairs ($p=0.001$), too
narrow a base to carry the aggregate, and not enough to offset the loss on
broad continuity.

\begin{figure}[!htb]
\centering
\includegraphics[width=0.92\linewidth]{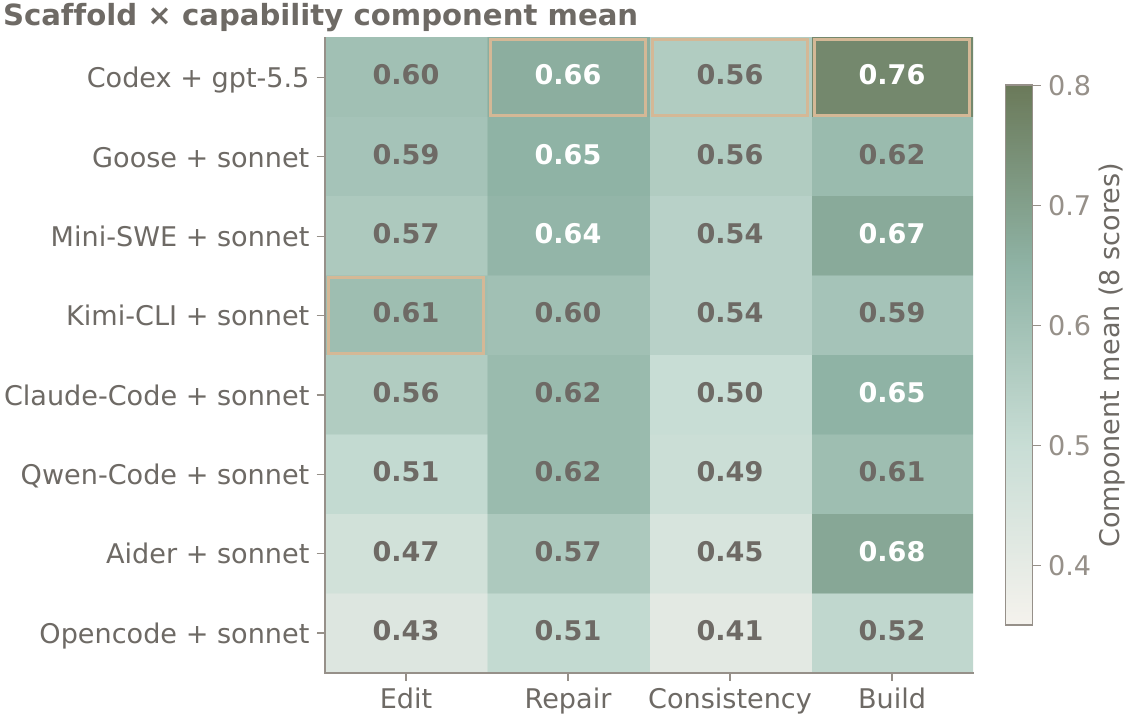}
\caption{Per-scaffold component mean by capability; orange box marks the
per-column winner. Codex with GPT-5.5 takes Build, Consistency and Repair,
Kimi-CLI takes Edit.}
\label{fig:appx_scaffold}
\end{figure}

\textbf{Robustness and validity ablations.} Table~\ref{tab:robustness} collects the single-variable checks
that could otherwise explain these results: edit format, vision on either side of
the loop, simulated-user wording, and rerun variability. No arm produces a net
strict-pass gain; the only arm whose net strict-pass count changes at all is
solver vision, at $-0.06$ percentage points, and the sim-user vision arm is the
only one where a single cell of thirteen moves in either direction; none clears the $|t|>2$ threshold on
$\Delta$Gate\%. No arm in this table moves the graded metric
outside the rerun noise floor; only the two prompt-scale interventions of
Section~\ref{sec:results} do, and even there the effect
is established at the gate level only; Appendix~\ref{app:regression} reports
the protected-state endpoint for every arm in this table and finds no significant
movement in any of them. Two of these rows exist specifically to rebut the
objection that the simulated user is doing the solver's diagnostic work: stripping
the persona down to raw failed-gate names and error strings moves $\Delta$Gate\%
by $-3.38$~pt ($t=-0.94$), and giving the simulated user a screenshot on
every turn moves it by $+2.43$~pt ($t=+0.70$). Neither is distinguishable
from zero, so the score is not an artifact of persona phrasing. For the two
arms that do move the gate rate, the movement is confined to gate quality:
truncating the solver's history shifts stage completion by only $-1.78$~pt
against its $-9.40$~pt on $\Delta$Gate\%, and leaves turn count flat.

\begin{table}[t]
\centering
\small
\caption{Single-variable robustness ablations. $\Delta$Gate\% is the paired mean shift in
19-gate pass rate over $n$ cells, normalised by its standard error as a paired
$t$. No row produces a net strict-pass gain.}
\label{tab:robustness}
\setlength{\tabcolsep}{3pt}
\begin{tabular}{lllrrrr}
\toprule
\rowcolor{HeaderColor}
\textbf{Ablation} & \textbf{Model \& leg} & \textbf{Direction} &
$n$ & $\Delta$\textbf{Gate\%} & $t$ & $\Delta$\textbf{Task\%} \\
\midrule
Edit-emission u-diff          & GPT-5.5 LLM       & vs JSON-patch      & 20 & $+3.16$ & $+0.62$ & $\ \ 0.00$ \\
Edit-emission full-file       & GPT-5.5 LLM       & vs JSON-patch      & 20 & $+4.74$ & $+1.26$ & $\ \ 0.00$ \\
Solver vision on/off          & 9-model dual-modal& VLM vs LLM         & 3{,}558 & $-0.28$ & $-0.87$ & $-0.06$ \\
Sim-user always vision        & Sonnet-4.6 LLM    & vs default         & 13 & $+2.43$ & $+0.70$ & $-7.69$ \\
Sim-user errmsg only          & Sonnet-4.6 VLM    & vs persona prose   & 14 & $-3.38$ & $-0.94$ & $\ \ 0.00$ \\
Rerun determinism             & 3-model, 5 slots  & 3 replays at $T=0$ & 15 & $\sigma{=}2.6$~gates & --- & $0/15$ flip \\
\bottomrule
\end{tabular}
\end{table}

\section{Discussion}
\label{app:discussion}

\textbf{The two verdicts come apart, and the gap is not small.} A target-only
metric is satisfied the moment the requested DOM appears. On our suite that
moment is routinely reached while the surrounding product is already broken:
28.2\% of the Change runs that clear every requested-work gate still fail a
conservative non-completion check, and among interactive runs whose target DOM
passes, 29.5\% still fail an interaction check on the very element they were
asked to produce. The first number says a correct target does not imply an
intact site; the second says it does not even imply a working feature. Both are
invisible to an endpoint DOM assertion. That is why the unit we score is the
change together with the website it lands in.

\textbf{Provenance is what makes a regression claim defensible.} Calling an
outcome a regression requires knowing the state was good before, and a
benchmark that only inspects the final workspace cannot know that. Our two
channels each carry that evidence by construction: one replays a state the run
itself cleared, the other checks a gate the untouched repository already
satisfied. The untouched-repository channel also shows that two of its gates
report one event: all 1{,}364 runs that leave a route blank also fail the
capture check on that route. Their union covers 18.3\% of the covered subset while they overlap on
2.7\%, so they are complementary views that partly coincide, not
one effect counted twice. We do not read the small overlap as independence:
under independence the expected overlap is about 6.4 runs against an observed
16. The replay channel then splits in a way no endpoint-only
harness can see: of the 10.8\% of completing runs that break a protected
state, a little over half finish with every conservative check clean, the
rest do not. Because the endpoint is a run-level verdict, not a
per-assertion repair record, that first group is invisible to a final-state
measure -- the run ends looking clean and the damage happened anyway.

\textbf{Long trajectories do not quietly fix themselves.} Within every fixed
stage-count bucket on Build the correlation between turns spent and stages
cleared is negative, median $-0.91$: units that clear finish quickly, while a
failed one consumes its full budget. A unit stops when it clears or when the
budget runs out, so the correlation partly reflects that rule; what it shows is
that turn count on its own does not separate a recovering trajectory from a
stuck one, which is a reason not to read persistence as progress. The failures are also not an artifact of
patch machinery. Of the 8,539 failing cells, 98.3\% land at least one edit and
only 1.7\% land none, so under two per cent of failures can be explained by an
agent that never managed to write to the workspace.

\textbf{More perception is not the missing ingredient.} On matched pairs
differing in whether the solver gets the screenshot feedback policy, the graded
score moves by $-0.0007$, indistinguishable from zero, and broad
visual-inclusive continuity moves against the screenshot arm. Assignment is not
the same as exposure: even among qualifying models, between 0.807 and 0.912 of
screenshot cells actually carry an image. The one facet that improves,
experience, is too narrow to carry the aggregate. We read this as a
statement about the bottleneck rather than about vision: on these tasks the most
likely constraint is knowing which prior behaviour is protected, which a
screenshot of the current page does not supply. No experiment here isolates that
hypothesis.

\textbf{The harness is part of the measurement.} In the full-spec scaffold arm,
the scaffold spread under a fixed backbone is larger than the effect of
swapping the backbone under a fixed scaffold, yet scaffold rank order is largely
preserved when the backbone changes. The two factors are close to additive: swapping the backbone helps
every scaffold, by between $0.60$ and $8.27$ points. The practical consequence
is that a frontend agent score is not interpretable without the scaffold that
produced it, and that the two can be improved separately.

\textbf{The simulated user is an evaluation interface, deliberately so.} One
capability-conditioned persona per task reveals requirements authored in advance and
gives grounded feedback, while the frozen verifier alone decides acceptance.
That separation is what makes runs comparable across thirteen models: the
conversation can vary without moving the bar. Multi-role interaction, adversarial
users and negotiated requirements are natural extensions, and because acceptance
is independent of the dialogue they can be added on the same frozen states and
gates without reauthoring the suite.

\section{Limitations and Scope}
\label{app:limitations}

Three boundaries define where the claims of Section~\ref{sec:results} apply.

\textbf{The twin result is stated for its regime.} The greenfield-twin finding,
that from-scratch construction scores below the matched in-place change on all
five legs, is established on 21 length-matched Change tasks whose twins are
mostly static or vanilla-node sites, because our authoring chain yields a blank
baseline only for those stacks. We state it for that regime, not for arbitrary
stacks.

\textbf{The turn budget is endogenous.} Failed units consume the stage cap while
cleared units stop early, so turn count is an outcome of the trajectory as much
as an input. We report persistence to the cap as an observation about how these
trajectories end, not a causal claim about what extra turns buy.

\textbf{Coverage is uneven and cells are clustered.} Provider availability
leaves coverage uneven, and several cells share a task. We state every
denominator, use matched pairs where one exists, report a
task-clustered interval beside each cell-level one, and read the pooled table
as observed coverage, the paired contrasts carrying the claims. The same applies to the
capability-against-regression comparison of Section~\ref{sec:results},
which thirteen models bound loosely but cannot rule out.

Two scope notes close the picture. Visual anchors are deterministic checks on
authored properties, not a measure of unrestricted aesthetic quality. And the
interventions of Section~\ref{sec:results} are prompt-scale probes aimed at
the graded gate metric; testing whether a prompt can move the rarer
protected-state endpoint needs a separately powered arm, as the arm-by-arm
figures in Appendix~\ref{app:regression} show.

\section{Reproducibility and Release Safeguards}
\label{app:reproducibility}

Each released task package contains repository provenance, allowed paths, the
ordered stage plan, route, viewport and action states, design anchors, deterministic
checks, reference metadata, and negative-control metadata. Evaluation starts from
a pristine workspace. Solver prompts and screenshots exclude reference patches,
bad solutions, and hidden gate implementation. Result rows carry task, model,
modality, specification timestamp, and trace identifiers; rows older than their
submission specification are excluded before analysis. The released manifest is
post-filter and carries the normalized labels, so both the exclusion and the
label mapping are materialized in the rows themselves.

\textbf{Gate judgement carries across environments.} The class-(a) baseline
(Appendix~\ref{app:regression}) is captured in the analysis environment rather
than on the evaluation cluster, so we calibrated the two against a known answer:
a reference implementation must pass every gate by admission. Re-capturing six
reference workspaces locally reproduces that verdict on the five that serve
offline: four pass all nineteen gates and one misses only
\texttt{visual\_regression}, the gate excluded throughout for its uncontrolled
scroll offset. The sixth does not start a server in the analysis harness and
yields no verdict either way. Every served recapture agrees with the cluster on
every nonvisual gate. Five workspaces are a small sample, so this supports rather
than establishes cross-environment equivalence for the eighteen gates the
class-(a) statistic uses; replaying the initial and the final workspace of the
same run side by side in one environment is the check that would settle it.

Ten of the baselines were captured later than the rest, on a rebuilt analysis
machine, which lets us separate two questions the first calibration conflates.
Repeating the known-answer test there gives the same result: of the reference
workspaces that serve, one passes seventeen of nineteen gates and three pass
eighteen, and the only gate all four miss is \texttt{visual\_regression}. The
environment therefore still judges a correct implementation correctly. Recapturing
the \emph{untouched} repository is a different matter: over thirteen slots and
247 gate verdicts, a recapture agrees with the frozen baseline on 85\% of them,
and the disagreements concentrate in \texttt{console\_errors},
\texttt{no\_horizontal\_overflow} and \texttt{visual\_anchor\_similarity}, in both
directions. A reference implementation sits far from those thresholds and a
partly broken repository sits on them, so the pristine side of the comparison is
the browser-version-sensitive one. We therefore read a single slot's baseline as
an environment-conditional reading rather than an absolute one; the ten late
captures move the union count by one event, from 106 to 107.

Repositories and local assets retain their original licenses. The task suite does
not require personal user data, production credentials, or deployment to public
services. Provider credentials, private billing metadata, and transient
workspaces are excluded from the release.

\section{Evidence-Based Case Studies}
\label{app:case_studies}

Cases (1) to (4) walk one live cell from each of the four \bench{}
capabilities, in the order the paper introduces them, and show what the frozen
contract catches on the corresponding real frontend. Cases (5) to (7) then take
the three contributions of Section~\ref{sec:intro} one at a time and show each
of them firing on a single task. Browser crops are actual screenshots retained
under \texttt{siflow/output\_b/cells/}; final checks and stage outcomes come
from \texttt{submission\_results.json}, \texttt{trace.json}, and
\texttt{result\_b.json}. All seven are chosen for the visual strength of the
reference UI and for a clean contract exercise, not as prevalence estimates.

Table~\ref{tab:case_map} maps the cases to the capability axes and what the
verifier grades. Case~(1) exercises a project-path selector that rewires the
visible set, Case~(2) tests preserved content on a rich album page,
Case~(3) checks a bound scroll-driven interaction survives a scene-composition
edit, and Case~(4) demands every unit clear within budget on a passing
final replay. The last three show the contract rejecting an
authored near-miss, protected-state replay firing on a run whose endpoint looks
clean, and the untouched-repository channel catching a property the run never
asserted.

\begin{table}[H]
\centering
\small
\setlength{\tabcolsep}{7pt}
\renewcommand{\arraystretch}{1.16}
\caption{Coverage of the seven case studies: the first four walk the four
capability axes and the last three walk the three contributions.}
\label{tab:case_map}
\begin{tabular}{@{}l l l l@{}}
\toprule
\rowcolor{HeaderColor}
\textbf{Case} & \textbf{Capability} & \textbf{Slot / model} & \textbf{What the verifier grades} \\
\midrule
(1) Path selector     & Edit        & 252~/ Hunyuan-3  & Interaction actually rewires the visible set \\
(2) Neon album        & Consistency & 020~/ Opus-4.8   & Preserved copy, artwork, ordered list \\
(3) Cinematic scroll  & Repair      & 178~/ Kimi-K3    & Cumulative replay of bound interactions \\
(4) Analytics board   & Build       & 211~/ Kimi-K3    & Every stage clears and the final suite passes \\
\midrule
(5) Authored near-miss & Consistency & 306~/ controls   & Admission, before any solver is run \\
(6) Dental practice   & Edit        & 114~/ Sonnet-4.6 & Protected-state replay over the trajectory \\
(7) Prompt explorer   & Repair      & 157~/ MiniMax-M3 & A gate the untouched repository passed \\
\bottomrule
\end{tabular}
\end{table}

\subsection{Case (1): A Stage Selector That Has to Actually Rewire the Page}

Slot 252 is an Edit task on \emph{Marlow Ferris}, a Vite + React + TypeScript
portfolio. A simulated PM asks, over four staged requirements, for the project
path selector to be clarified (s1), carried consistently onto the portfolio
view (s2), \emph{genuinely wired up} on desktop rather than merely looking
selected (s3), and kept inside a narrow viewport (s4); two reworks, a late
wording and ordering change and a mobile QA bug, then verify everything again
already cleared. Hunyuan-3-Preview drove the screenshot leg for 39 turns,
asked 11 clarifying questions, and applied 15 edits.

\begin{figure}[!htb]
  \centering
  \includegraphics[width=\textwidth]{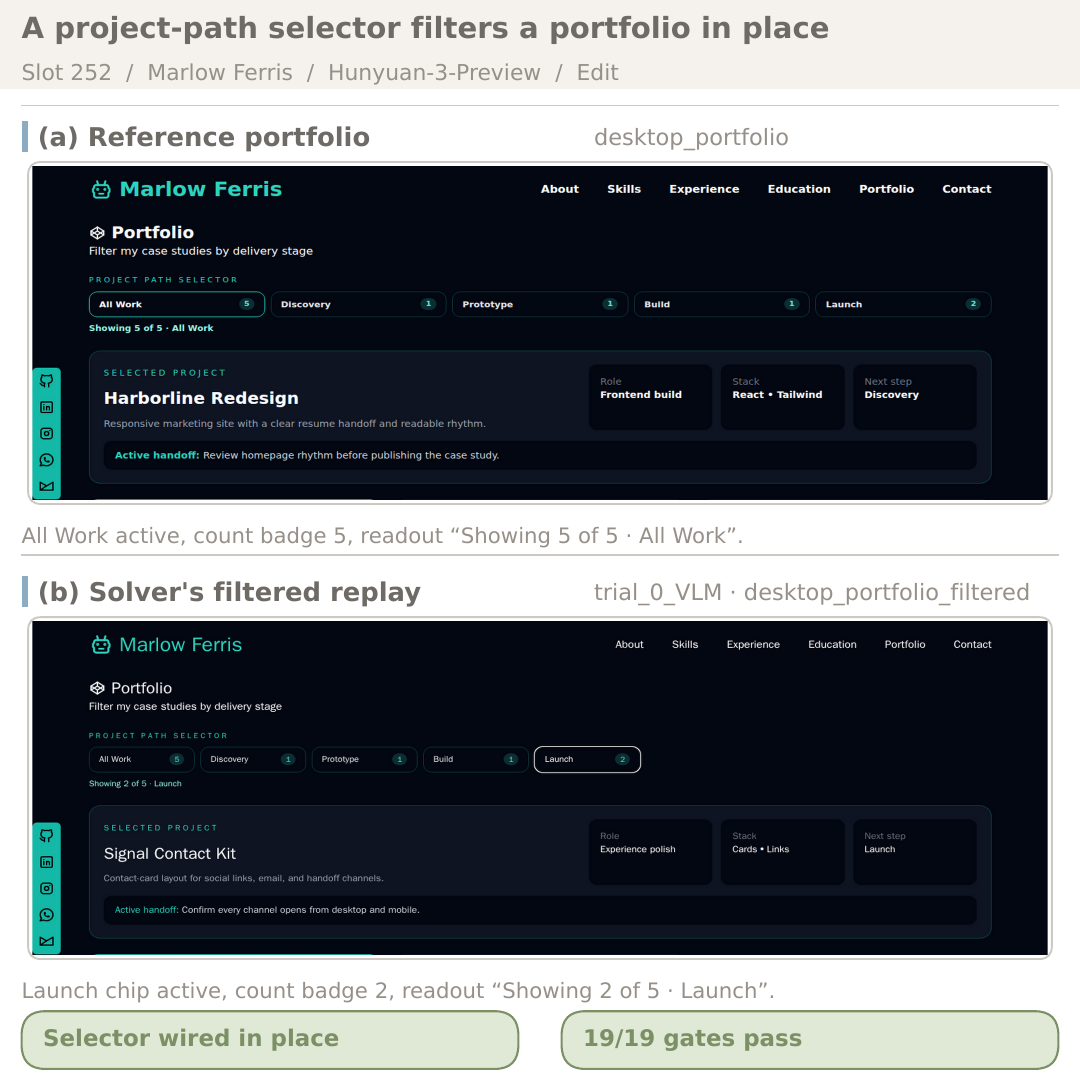}
  \caption{\textbf{Edit capability, project path selector (slot 252).}
  (a) Reference portfolio: \emph{All Work} active, count badge 5, readout
  \emph{Showing 5 of 5 $\cdot$ All Work}. (b) Hunyuan-3-Preview's replay after
  a real click on \emph{Launch}: count badge 2, readout \emph{Showing 2 of 5
  $\cdot$ Launch}, and the selected-project panel switches with it. The stage-s3
  gate requires all three at once: a genuine interaction, a state that
  differs from its predecessor, and the exact live count.}
  \label{fig:case_pretty_portfolio}
\end{figure}

Figure~\ref{fig:case_pretty_portfolio} shows stage s3, the stage that makes the
capability nontrivial. Its checkpoint is a single
\texttt{interaction\_assertions} clause: the state
\texttt{desktop\_portfolio\_filtered} must be reached from
\texttt{desktop\_portfolio} by a real click
(\texttt{require\_interaction: true}), it must differ from the state it came
from (\texttt{changed\_from}), and the resulting DOM must contain the exact
string \emph{Showing 2 of 5}. A chip that merely receives an active outline
satisfies none of the three. In the replay, clicking \emph{Launch} switches the
active chip, drops the readout from \emph{Showing 5 of 5 $\cdot$ All Work} to
\emph{Showing 2 of 5 $\cdot$ Launch}, and swaps the selected-project panel from
\emph{Harborline Redesign} to \emph{Signal Contact Kit}: the count, the
label and the content all move together.

The cell clears s2, s3 and s4 and passes all 19 final gates with zero
protected-state regressions, but it never clears s1, and neither rework is
cleared; its long-horizon score is therefore $0.875$, not $1$.

\textbf{What this establishes.} Edit-family tasks in \bench{} do not grade the
requested DOM addition on its own; they grade whether the addition is
\emph{live}. The frozen contract for s3 names the predecessor state, demands
that the successor be reached by an interaction, and pins the exact
\emph{N of M} string that interaction must produce, so a purely cosmetic chip
row fails the stage while still rendering perfectly. Because the two reworks
verify every earlier stage again, the same contract also asks whether a late
wording change and a mobile fix can be absorbed while everything earlier keeps
passing. Here the earlier states do hold (the run records no protected-state
regression), but neither rework clears, which is why it stops at $0.875$
despite a clean final gate sheet. That gap is the common shape of a strong \bench{} run:
the visible artefact looks finished, and the staged contract still records what
was left behind.

\subsection{Case (2): Neon Album Continuity Across a Consistency Task}

Slot 020 asks Claude Opus 4.8 (VLM) to extend a record-label site while
retaining the frozen \emph{Neon Dreams} album page. The reference layout is a
cyberpunk-cityscape artwork hero, a strictly ordered ten-track list, an
artist byline, and a now spinning rail at the bottom. The requested change is
to append a \emph{next drop} teaser into the now spinning rail without
altering any incumbent element. In the solver's replay every one of the four
protected regions survives the diff: the artwork tile, all ten track lines
in their original order, the \emph{by Echo Lume} byline, and the release
meta block still read identically to the reference; the now spinning rail
gains a valid \emph{Wanderer's Rest} tile in the new slot.

\begin{figure}[!htb]
  \centering
  \includegraphics[width=\textwidth]{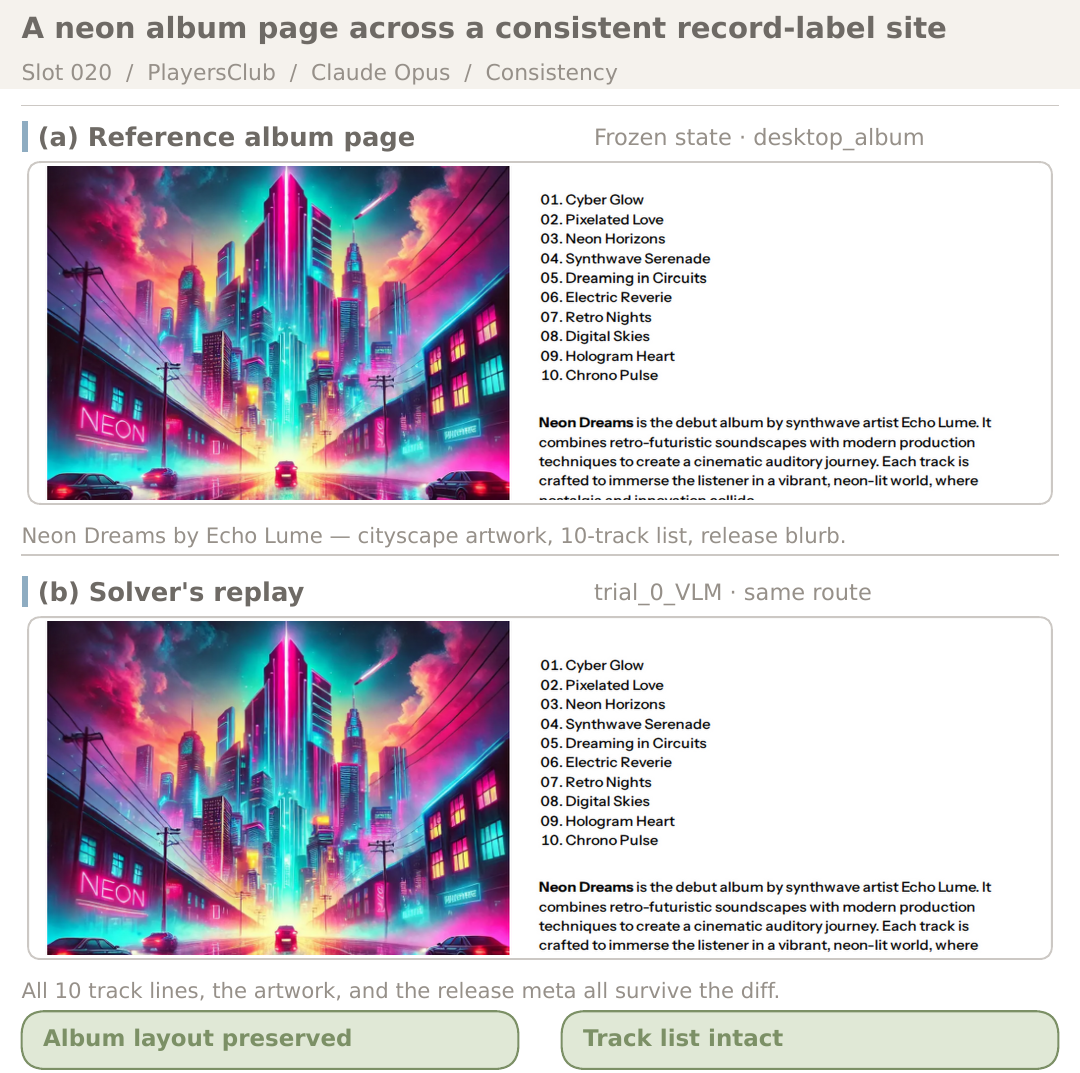}
  \caption{\textbf{Consistency preserved on a visually rich album page (slot 020).}
  Both panels are cropped to the same region of the same route.
  (a) Reference album page in its frozen state: the cyberpunk artwork hero, all
  ten ordered tracks, and the release blurb. (b) The solver's replay: artwork,
  track order, and blurb match the reference after the edit; the panels compare
  the rendered region, and byte-level identity is not what is shown here.
  The byline and the now spinning rail sit below this crop and are verified by
  the gates rather than shown here.}
  \label{fig:case_pretty_synthwave}
\end{figure}

Figure~\ref{fig:case_pretty_synthwave} places the two renders side by side.

\textbf{What this establishes.} Consistency-family tasks in \bench{} do not
grade on the requested feature alone; they grade on whether the preexisting
site (routes, layouts, protected copy, artwork tiles, ordered list content)
survives every diff. In slot 020 the frozen contract explicitly names the
ten track lines (in order), the album masthead, the artwork tile, and the
release meta as \emph{preserved-content} anchors; the verifier crops all
four regions on the solver's final replay and compares them to the frozen
reference. It also fires the now spinning rail's next-drop cue again to confirm
that the newly added tile is a valid drop and not a placeholder. This case is
instructive precisely because a naive completion grader would give the same
solver full credit on a much narrower requested change (adding a next-drop
rail); only the Consistency contract catches the much larger question of
whether the ten-track list, artwork, and release meta all made it through
unchanged. Here the answer is yes end-to-end, and the retained interaction
on the now spinning rail confirms it.

\subsection{Case (3): A Cinematic 3D Scroll Survives Revalidation}

Slot 178 asks Kimi-K3 (LLM) to repair a Codrops cinematic scene director on
a GSAP + OGL Astro site. The reference \texttt{desktop\_home} carries a
twelve-frame cylindrical viewport driven by scroll, a \emph{scene director}
control panel (four scene chips: All chapters, Opening, Ascent, Skyline) and
a \emph{Now cueing} status line. Editing the panel to add a second demo
variant is a scene-composition change: it must render a new scroll-linked
scene without breaking any previously cleared chapter cue or the
runtime metadata (frames, chapters, runtime, motion). In the solver's replay
the new \emph{demo-2} route renders a black \emph{DISCOVER --- The Future of
Architecture} scene, the ten chapter chips (DISCOVER, CYBERPUNK, PENETRATION,
ELEVATION, INTERLOCK, VISION, ARCHITECTURE, FUTURE, SUMMIT, WELCOME) still
cue the correct scroll progress, and the scene-director's field-notes
rail stays bound to the active scene.

\begin{figure}[!htb]
  \centering
  \includegraphics[width=\textwidth]{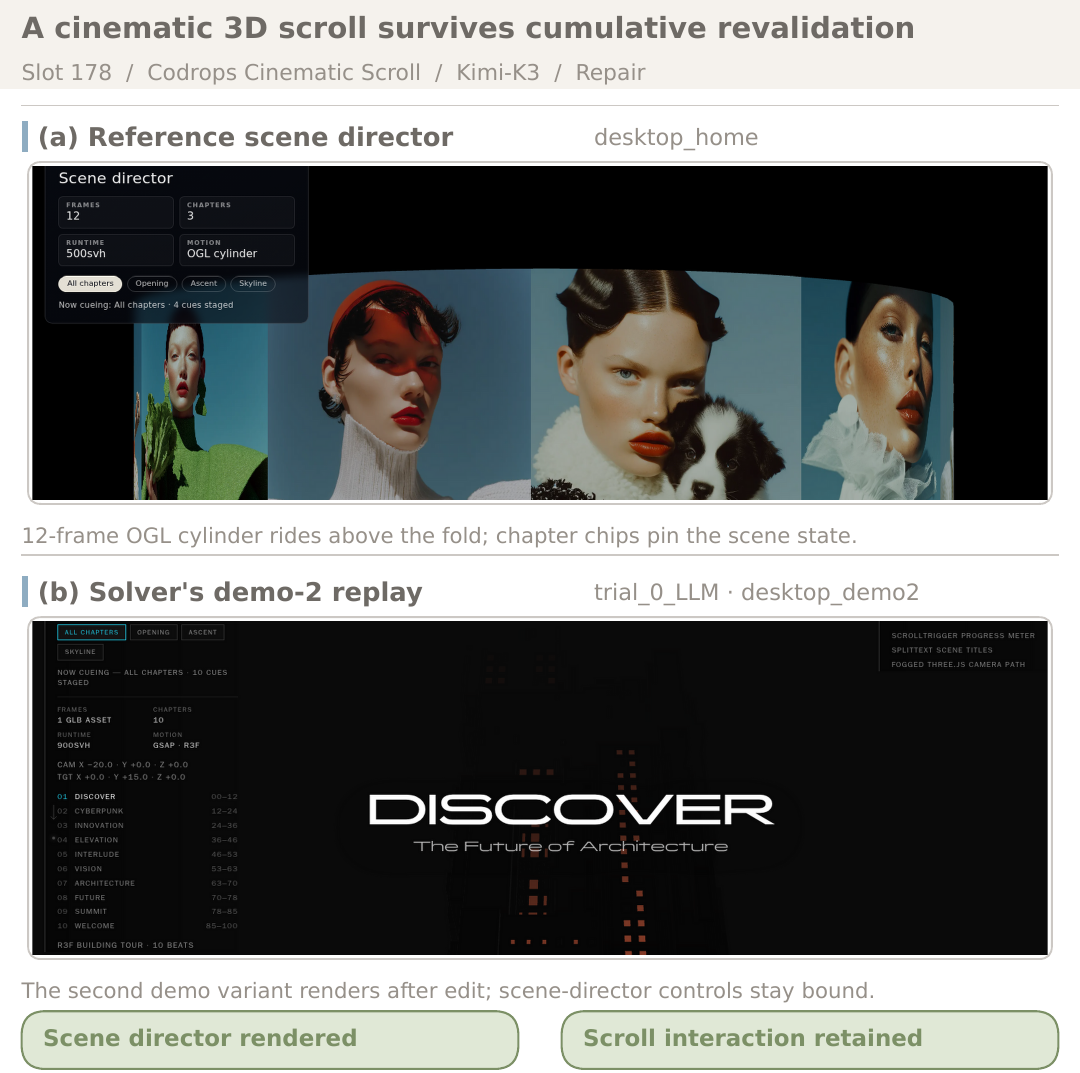}
  \caption{\textbf{A cinematic 3D scroll survives cumulative revalidation (slot 178).}
  (a) Reference home cylinder with the scene-director control open. (b) The
  solver's demo-2 replay: the second scene variant renders and the
  scroll-driven interactions on the director stay bound.}
  \label{fig:case_pretty_cinematic}
\end{figure}

Figure~\ref{fig:case_pretty_cinematic} shows the reference scene and the replay.

\textbf{What this establishes.} Repair-family tasks in \bench{} check for a
much stricter property than a green diff: after the edit lands, every
previously cleared interaction on the affected \emph{and neighbouring} routes
must still fire correctly at final replay. Slot 178's frozen contract binds
the scene director to a twelve-frame OGL cylinder with two demo variants, a
ten-chip chapter navigator, and four runtime-metadata fields. The verifier
runs the scroll interaction again on \texttt{desktop\_demo2} after the edit,
reads the \emph{SCENE 01/10 DISCOVER} state again at the first stop, cycles
through the remaining chapter chips, and confirms that (a) the cylinder
still transitions across all twelve frames without missing a chapter cue,
(b) the runtime line still reads \emph{500svh}, and (c) neither Demo~1 nor
the tutorial route regresses. Nothing about the diff itself would tell us
whether the scroll-linked cues survived; only cumulative replay does. The
result is meaningful because scene-composition edits (adding the demo-2
variant, retitling chapter cues, threading a new motion mode) routinely
break the scroll bindings in similar Codrops-based sites; here the solver
preserves them.

\subsection{Case (4): An Analytics Dashboard Passes Every Checkpoint}

\begin{figure}[!htb]
  \centering
  \includegraphics[width=\textwidth]{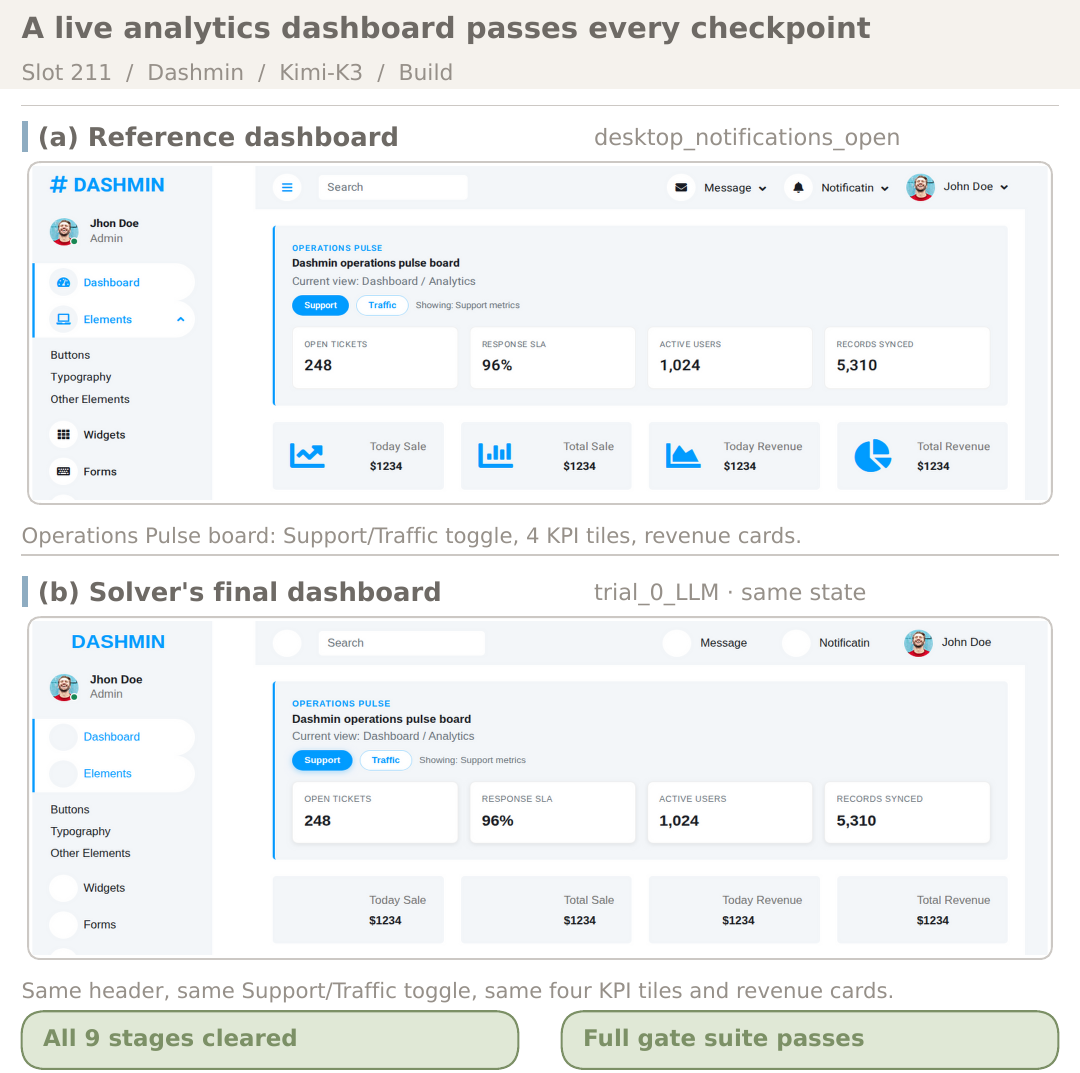}
  \caption{\textbf{A live analytics dashboard passes every checkpoint (slot 211).}
  (a) Reference dashboard at \texttt{desktop\_notifications\_open}. (b) The
  solver's final dashboard on the same state: identical KPI values and chart
  series, notifications rail open. Nine stages cleared, full gate suite passed.}
  \label{fig:case_pretty_dashboard}
\end{figure}

Slot 211 asks Kimi-K3 (LLM) to build a full \emph{Operations Pulse}
admin board on top of the Dashmin shell over nine ordered stage/rework
units: shell scaffolding, sidebar navigation, the pulse-board header, the
four KPI tiles (\emph{Open tickets}, \emph{Response SLA}, \emph{Active
users}, \emph{Records synced}), the Support-versus-Traffic toggle, the two
chart cards (\emph{Worldwide sales}, \emph{Sales \& Revenue}), the Recent
Sales table, the notifications rail, and a mobile-containment pass at 390
pixels. Nine of nine stage/rework units clear inside their 12-turn budgets,
and the final replay passes the nineteen-gate suite. The reference and solver
captures at \texttt{desktop\_notifications\_open} both show the identical four
KPI tiles, the Support/Traffic toggle in the same
state, the two chart cards with the same series, and the notifications rail
open at the same items (Figure~\ref{fig:case_pretty_dashboard}).

\textbf{What this establishes.} Build-family tasks are the strictest signal
\bench{} produces: a run is only marked passed when \emph{every}
stage/rework unit clears within its 12-turn budget \emph{and} the final
replay passes all nineteen gates on the retained state. In slot 211 both
hold: each KPI tile renders with the frozen value, the Support/Traffic
toggle switches its metric label without breaking the underlying series,
the two chart cards read the same USA/UK/AU distribution as the reference,
and the notifications rail opens on demand. The trajectory itself is also
clean: no stage was capped, no rework requested by the PM unit went
unresolved, and the mobile-containment gate at a 390-pixel viewport passes
on the last checkpoint. A positive Build signal therefore asserts not just a
passing final render, but a passing final render on top of nine successful
budgeted milestones.

\textbf{Taken together}, the four capability cases rest on different evidence:
Edit on a shared component and its cross-route reuse, Consistency on preserved
copy, artwork and ordered lists, Repair on cumulative replay of cleared
interactions after a heavy edit, and Build on every unit clearing inside its
turn budget. A single \emph{pass} therefore carries different weight by
capability, which is why \bench{} keeps three records: the final
artifact, what it preserves, and which stages cleared inside the budget. The three cases that follow take the
contributions one at a time on the same evidence.

\subsection{Case (5): The Contract Rejects an Authored Near-Miss}

\begin{figure}[!htb]
  \centering
  \includegraphics[width=0.80\textwidth]{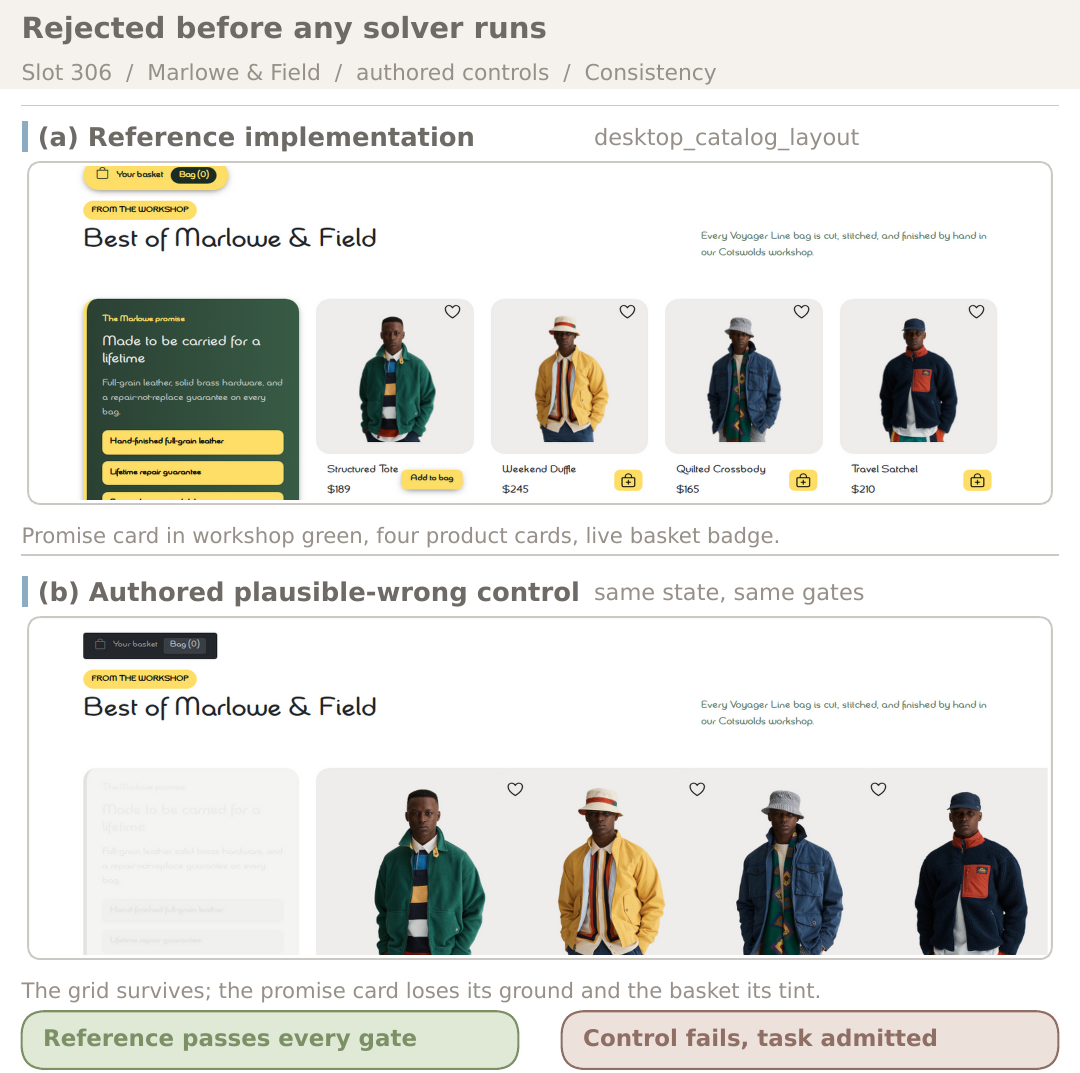}
  \caption{\textbf{Rejected before any solver runs (slot 306).} (a) The
  reference implementation at \texttt{desktop\_catalog\_layout}. (b) The
  authored plausible-wrong control at the same state, under the same gates. The
  product grid survives; the promise card and basket badge lose their
  treatment, and the contract rejects it.}
  \label{fig:case_pretty_contract}
\end{figure}

Slot 306 is a \emph{Marlowe \& Field} storefront whose Consistency task asks
for a workshop-promise card beside the product grid. Admission runs twice on
this package before any solver sees it. The reference patch is applied to the
initial repository and must clear every gate; an authored plausible-wrong
solution is applied to the same repository and must fail at least one.
Figure~\ref{fig:case_pretty_contract} shows both at
\texttt{desktop\_catalog\_layout}. The control is not a strawman: it keeps the
four product cards, the prices, the wishlist affordances and the heading, and a
reader skimming the page would take it for a working implementation. What it
drops is the promise card's ground and the basket badge's tint, exactly the
class of change the task is about, and the frozen gates reject it.

\textbf{What this establishes.} Because the reference passed and the control
failed on this package, a later model failure here cannot be explained by an
impossible requirement or by a gate that waves a plausible regression through.
This held for all 479 references and all 479 controls, which is what makes the
numbers in Section~\ref{sec:experiments} readable as model behaviour.

\subsection{Case (6): A Clean Endpoint on Top of a Damaged Trajectory}

\begin{figure}[!htb]
  \centering
  \includegraphics[width=0.80\textwidth]{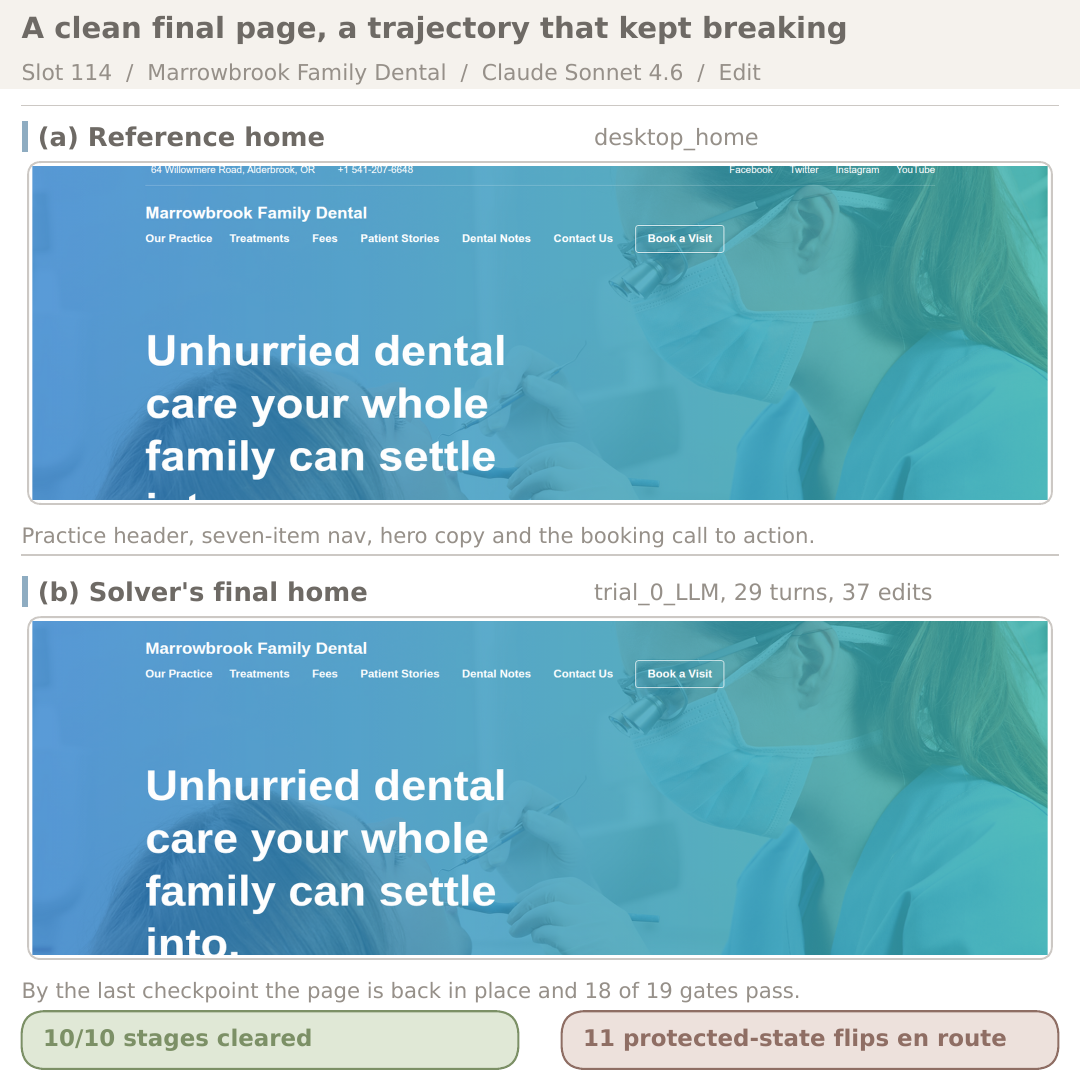}
  \caption{\textbf{A clean final page on a trajectory that kept breaking (slot
  114).} (a) The reference home at \texttt{desktop\_home}. (b) The solver's
  final home at the same state after 29 turns and 37 edits. Ten of ten stages
  cleared and 18 of 19 gates pass, the one failure being the visual check.
  Neither image can show what the cumulative record holds: eleven
  protected-state flips along the way.}
  \label{fig:case_pretty_dental}
\end{figure}

Slot 114 is the \emph{Marrowbrook Family Dental} practice site, and its Edit
task runs ten ordered units across the home, treatments and pricing routes.
Claude Sonnet 4.6 clears all ten of them over 29 turns and 37 applied edits,
and the final workspace passes eighteen of nineteen gates.
Figure~\ref{fig:case_pretty_dental} shows why an endpoint-only harness would
record this as a near-perfect run: the delivered home page is the reference
page. Protected-state replay tells a different story about how it got there.
Eleven times during the trajectory, an assertion that had already cleared at an
earlier stage failed at a later checkpoint, and each of those flips was repaired
before the run ended, which is why the endpoint gate sheet shows no trace of a single one of them.

\textbf{What this establishes.} This cell sits in the 38 of 674 completing runs
that regress a protected state and still finish with every conservative check
clean, the row of Table~\ref{tab:regression_endpoints_full} only cumulative
verification can populate. The clean endpoint and the eleven flips are both
recorded truthfully and they answer different questions, but a benchmark that
keeps only the endpoint cannot tell a steady run from a thrashing one that
happened to land.

\subsection{Case (7): A Property the Untouched Repository Already Had}

\begin{figure}[!htb]
  \centering
  \includegraphics[width=0.80\textwidth]{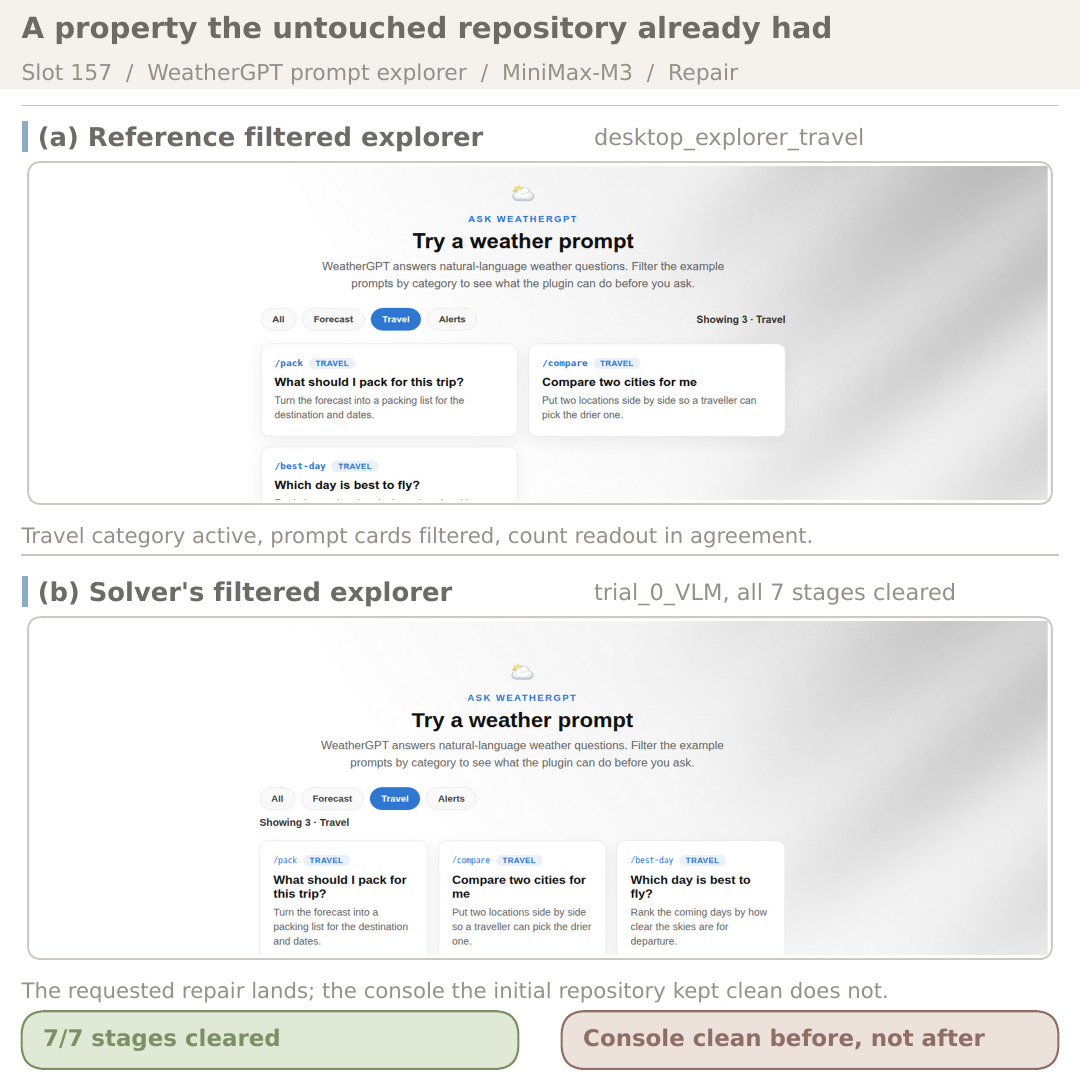}
  \caption{\textbf{A property the untouched repository already had (slot 157).}
  (a) The reference filtered explorer at \texttt{desktop\_explorer\_travel}.
  (b) The solver's replay of the same state: same active chip, same three
  travel prompts, same count readout. Seven of seven stages cleared; the
  console failure is one the second channel records and no image can show.}
  \label{fig:case_pretty_weather}
\end{figure}

Slot 157 is the \emph{WeatherGPT} prompt explorer, and its Repair task asks for
the category filter to apply correctly to the prompt cards.
Figure~\ref{fig:case_pretty_weather} shows the reference and the MiniMax-M3
replay at \texttt{desktop\_explorer\_travel}: the Travel chip is active in both,
the same three travel prompts are listed, and the readout reads
\emph{Showing 3 $\cdot$ Travel} in both. All seven stage units cleared and every
completion gate passes, so the requested repair landed. Protected-state replay
records nothing, because no stage in this run ever asserted the browser console.
The untouched-repository channel does: the repository as handed to the solver
serves this route with a clean console, and the delivered workspace no longer
does once the repair has landed.

\textbf{What this establishes.} This is a class-(a) regression, invisible to
every channel except the second one: target-only scoring passes it because the
requested repair is present, and replay cannot speak to it because it protects
only what the run itself established. Only a gate suite run against the
repository as handed to the solver can say that this property was working
before and is not working now.


\end{document}